\documentclass[10pt,letterpaper]{article}
\usepackage[top=0.85in,left=2.75in,footskip=0.75in]{geometry}

\usepackage{amsmath,amssymb}

\usepackage{changepage}

\usepackage[utf8x]{inputenc}

\usepackage{textcomp,marvosym}

\usepackage{cite}

\usepackage{nameref,hyperref}
\hypersetup{
    colorlinks=true,
    linkcolor=blue,
    filecolor=magenta,      
    urlcolor=blue,
    pdfpagemode=FullScreen,
}

\usepackage[right]{lineno}

\usepackage{microtype}
\DisableLigatures[f]{encoding = *, family = * }

\usepackage[table]{xcolor}[]

\usepackage{array}

\newcolumntype{+}{!{\vrule width 2pt}}

\newlength\savedwidth

\raggedright
\usepackage[aboveskip=1pt,labelfont=bf,labelsep=period,justification=raggedright,singlelinecheck=off]{caption}

\makeatletter
\renewcommand{\@biblabel}[1]{\quad#1.}
\makeatother

\usepackage{lastpage,fancyhdr,graphicx}
\usepackage{epstopdf}
\fancyheadoffset[L]{2.25in}
\fancyfootoffset[L]{2.25in}
\usepackage{bm}
\usepackage{mathtools}
\usepackage[inline]{enumitem}

\usepackage{tikz,lmodern}
\usepackage[most]{tcolorbox}
\usepackage[note-field=title]{notes2bib}
\usepackage{caption}

\newcommand*{\argmin}{\operatornamewithlimits{argmin}\limits}
\newcommand*{\argmax}{\operatornamewithlimits{argmax}\limits}
\newcommand{\powerset}{\mathbb{P}}
\newcommand{\D}{D}

\newcommand{\T}{\mathcal{T}}

\newcommand{\p}{\theta} 
\newcommand{\Part}{\Theta}
\renewcommand{\O}{\emptyset}

\newcommand{\ce}{e}

\newcommand{\inp}{\leftarrow}
\newcommand{\out}{\rightarrow}
\newcommand{\both}{\leftrightarrow}
\renewcommand{\sp}{\bar{s}}
\newcommand{\sm}{\bar{s}}
\renewcommand{\sc}{s}
\newcommand{\up}{\bar{u}}
\newcommand{\uc}{u}

\DeclareMathOperator{\ii}{\mathit{ii}} 

\begin{document}
\vspace*{0.2in}

\begin{flushleft}
{\Large
\textbf\newline{Integrated information theory (IIT) 4.0: Formulating the properties of phenomenal existence in physical terms} 
}
\newline
\\
Larissa Albantakis\textsuperscript{1\Yinyang},
Leonardo Barbosa\textsuperscript{1,2\Yinyang},
Graham Findlay\textsuperscript{1,3\Yinyang},
Matteo Grasso\textsuperscript{1\Yinyang},
Andrew M Haun\textsuperscript{1\Yinyang},
William Marshall\textsuperscript{1,4\Yinyang},
William GP Mayner\textsuperscript{1,3\Yinyang},
Alireza Zaeemzadeh\textsuperscript{1\Yinyang},
Melanie Boly\textsuperscript{1,5},
Bjørn E Juel\textsuperscript{1,6},
Shuntaro Sasai\textsuperscript{1,7},
Keiko Fujii\textsuperscript{1},
Isaac David\textsuperscript{1},
Jeremiah Hendren\textsuperscript{1,8},
Jonathan P Lang\textsuperscript{1},
Giulio Tononi\textsuperscript{1*}
\\
\bigskip
\textbf{1} Department of Psychiatry, University of Wisconsin, Madison, WI 53719, USA
\\
\textbf{2} 
Fralin Biomedical Research Institute at VTC, Virginia Tech, Roanoke, VA 24016, USA
\\
\textbf{3} Neuroscience Training Program, University of Wisconsin, Madison, WI 53705, USA
\\
\textbf{4} Department of Mathematics and Statistics, Brock University, St. Catharines, ON L2S 3A1, Canada 
\\
\textbf{5} Department of Neurology, University of Wisconsin, Madison, WI 53719, USA
\\
\textbf{6}  Institute of Basic Medical Sciences, University of Oslo, Oslo, 0372, Norway
\\
\textbf{7} Araya Inc., Tokyo, 107-0052, Japan
\\
\textbf{8} Graduate School Language \& Literature, Ludwig Maximilian University of Munich, Munich, 80799, Germany
\bigskip

%
%
\Yinyang These authors contributed equally to this work.



* gtononi@wisc.edu

\end{flushleft}
\section*{Abstract}
This paper presents Integrated Information Theory (IIT) 4.0. IIT aims to account for the properties of experience in physical (operational) terms. It identifies the essential properties of experience (axioms), infers the necessary and sufficient properties that its substrate must satisfy (postulates), and expresses them in mathematical terms. In principle, the postulates can be applied to any system of units in a state to determine whether it is conscious, to what degree, and in what way. IIT offers a parsimonious explanation of empirical evidence, makes testable predictions, and permits inferences and extrapolations. IIT 4.0 incorporates several developments of the past ten years, including a more accurate translation of axioms into postulates and mathematical expressions, the introduction of a unique measure of intrinsic information that is consistent with the postulates, and an explicit assessment of causal relations. By fully unfolding a system's irreducible cause--effect power, the distinctions and relations specified by a substrate can account for the quality of experience.


\section*{Author summary}
IIT aims to account for consciousness and its properties in physical terms. The theory identifies the essential properties of experience (axioms), formulates them as physical properties in terms of cause--effect power (postulates), and provides a mathematical formalism for assessing those properties. This formalism can be employed to ``unfold'' a cause--effect structure from a substrate constituted of units in a state, whose interactions can be characterized as interventional conditional probabilities. According to IIT, all the properties of an experience can be accounted for, in physical terms, by those of a cause--effect structure that satisfies its postulates. The theory is consistent with neurological data and has led to successful experimental predictions. As the latest iteration of the theory, IIT 4.0 incorporates several developments pursued over the past ten years.


\section*{Introduction}

A scientific theory of consciousness should account for experience, which is subjective, in objective terms \cite{Ellia2021}. Being conscious---having an experience---is understood to mean that ``there is something it is like to be'' \cite{Nagel1974}: something it is like to see a blue sky, hear the ocean roar, dream of a friend’s face, imagine a melody flow, contemplate a choice, or reflect on the experience one is having.

IIT aims to account for phenomenal properties---the properties of experience---in physical terms. IIT's starting point is experience itself rather than its behavioral, functional, or neural correlates \cite{Ellia2021}. Furthermore, in IIT ``physical'' is meant in a strictly operational sense---in terms of what can be observed and manipulated.

The starting point of IIT is the existence of an experience, which is immediate and irrefutable \cite{Tononi2015_scholarpedia, Tononi_OnBeing}. From this ``zeroth'' axiom, IIT  sets out to identify the essential properties of consciousness---those that are immediate and irrefutably true of every conceivable experience. These are IIT's five axioms of phenomenal existence: every experience is for the experiencer (intrinsicality), specific (information), unitary (integration), definite (exclusion), and structured (composition).

Unlike phenomenal existence, which is immediate and irrefutable (an axiom), physical existence is an explanatory construct (a postulate) and it is assessed operationally from within consciousness: in physical terms, to be is to have cause--effect power (see Box 2: Principle of being). In other words, something can be said to exist physically if it can ``take and make a difference''---bear a cause and produce an effect---as judged by a conscious observer/manipulator.

The next step of IIT is to formulate the essential phenomenal properties (the axioms) in terms of corresponding physical properties (the postulates). This formulation is an ``inference to a good explanation'' and rests on basic assumptions such as realism, physicalism, and atomism (see Box 1: Methodological guidelines of IIT). If IIT is correct, the substrate \bibnote{A substrate should be understood as a set of units that can be observed and manipulated}
of consciousness, beyond having cause--effect power (existence), must satisfy all five essential phenomenal properties in physical terms: its cause--effect power must be for itself (intrinsicality), specific (information), unitary (integration), definite (exclusion), and structured (composition).

On this basis, IIT proposes a fundamental explanatory identity: an experience is identical to the cause--effect structure unfolded from a maximal substrate (defined below). Accordingly, all the specific phenomenal properties of any experience must have a good explanation in terms of the specific physical properties of the corresponding cause--effect structure, with no additional ingredients.

IIT formulates the postulates in a mathematical framework that is in principle applicable to general models of interacting units (but see \bibnote{As mentioned in the section ``Determining maximal unit grains,'' a substrate unit must be maximally irreducible within, which is likely the case for ``real'' neurons in the brain, but is not the case for ``virtual,'' simulated neurons in a computer program}). A mathematical framework is needed (i) to evaluate whether the theory is self-consistent and compatible with our overall knowledge about the world, (ii) to make specific predictions regarding the quality and quantity of our experiences and their substrate within the brain, and (iii) to extrapolate from our own consciousness to infer the presence (or absence) and nature of consciousness in beings different from ourselves.

Ultimately, the theory should account for why our consciousness depends on certain portions of the  world and their state, such as certain regions of the brain and not others, and for why it fades during dreamless sleep, even though the brain remains active. 
It should also account for why an experience feels the way it does---why the sky feels extended, why a melody feels flowing in time, and so on. 
Moreover, the theory makes several predictions concerning both the presence and the quality of experience, some of which have been and are being tested empirically \cite{Tononi2016}.

While the main tenets of the theory have remained the same, its formal framework has been progressively refined and extended \cite{Tononi2003, Tononi2004, Balduzzi2008, Oizumi2014}. 
Compared to IIT 1.0 \cite{Tononi2003, Tononi2004}, 2.0 \cite{Balduzzi2008, Balduzzi2009}, and 3.0 \cite{Oizumi2014}, IIT 4.0 presents a more complete, self-consistent formulation and incorporates several recent advances \cite{Albantakis2019, Haun2019, Barbosa2021, Marshall2022}. Chief among them are a more accurate translation of the axioms into postulates and mathematical expressions, the introduction of an Intrinsic Difference (ID) measure \cite{Barbosa2020, Barbosa2021} that is uniquely consistent with IIT's postulates, and the explicit assessment of causal relations \cite{Haun2019}.

In what follows, after introducing IIT's axioms and postulates, we provide its updated mathematical formalism. In the ``Results and discussion'' section, we apply the mathematical framework of IIT to representative examples and discuss some of their implications. 
The article is meant as a reference for the theory's mathematical formalism, a concise demonstration of its internal consistency, and an illustration of how a substrate's cause--effect structure is unfolded computationally. 
A discussion of the theory's motivation, its axioms and postulates, and its assumptions and implications can be found in a forthcoming book \cite{Tononi_OnBeing} and wiki \cite{IITWiki} as well as in several publications \cite{Albantakis2020, Ellia2021, Grasso2021, Tononi2022, Tononi2015, Findlay2019, AlbantakisQIIT}. A survey of the explanatory power and experimental predictions of IIT can be found in \cite{Tononi2016}. 
The way IIT's analysis of cause--effect power can be applied to actual causation, or ``what caused what,'' is presented in \cite{Albantakis2019}.

\section*{From phenomenal axioms to physical postulates}

\subsection*{Axioms of phenomenal existence}
That experience exists---that ``there is something it is like to be''—is immediate and irrefutable, as everybody can confirm, say, upon awakening from dreamless sleep. Phenomenal existence is immediate in the sense that my experience is simply there, directly rather than indirectly: I do not need to infer its existence from something else. It is irrefutable because the very doubting that my experience exists is itself an experience that exists---the experience of doubting \cite{Ellia2021, Tononi2015_scholarpedia}. Thus, to claim that my experience does not exist is self-contradictory or absurd. 
The existence of experience is IIT's zeroth axiom.

\begin{description}
    \item[Existence] Experience \textit{exists}: there is \textit{something}.
\end{description}

Traditionally, an axiom is a statement that is assumed to be true, cannot be inferred from any other statement, and can serve as a starting point for inferences. 
The existence of experience is the ultimate axiom---the starting point for everything, including logic and physics.

On this basis, IIT proceeds by considering whether experience---phenomenal existence---has some axiomatic or essential properties, properties that are immediate and irrefutably true of every conceivable experience. Drawing on introspection and reason, IIT identifies the following five:

\begin{description}
    \item[Intrinsicality] Experience is \textit{intrinsic}: it exists \textit{for itself}.
    \item[Information] Experience is \textit{specific}: it is \textit{the way it is}.
    \item[Integration] Experience is \textit{unitary}: it is \textit{a whole}, \textit{irreducible} to separate experiences.
    \item[Exclusion] Experience is \textit{definite}: it is \textit{this} whole.
    \item[Composition] Experience is \textit{structured}: it is composed of \textit{distinctions} and the \textit{relations} that bind them together, yielding a \textit{phenomenal structure}.
\end{description}

To exemplify, if I awaken from dreamless sleep, and experience the white wall of my room, my bed, and my body, the experience not only exists, immediately and irrefutably, but 1) it exists for me, 2) it is specific (the wall is a wall and it is white), 3) it is unitary (the left side is not experienced separately from the right side, and vice versa), 4) it is definite (it includes the visual scene in front of me—neither less, say, its left side only, nor more, say, the wall behind my head), 5) it is structured by distinctions (the wall, the bed, the body) and relations (the body is on the bed, the bed in the room).

The axioms are not only immediately given, but they are irrefutably true of every conceivable experience. For example, once properly understood, the unity of experience cannot be refuted. Trying to conceive of an experience that were not unitary leads to conceiving of two separate experiences, each of which is unitary, which reaffirms the validity of the axiom. Even though each of the axioms spells out an essential property in its own right, the axioms must be considered together to properly characterize phenomenal existence. 

IIT takes the above set of axioms to be complete: there are no further properties of experience that are essential. Other properties that might be considered as candidates for axiomatic status include space (experience typically takes place in some spatial frame), time (an experience usually feels like it flows from a past to a future), change (an experience usually transitions or flows into another), subject--object distinction (an experience seems to involve both a subject and an object), intentionality (experiences usually refer to something in the world, or at least to something other than the subject), a sense of self (many experiences include a reference to one’s body or even to one’s narrative self), figure--ground segregation (an experience usually includes some object and some background), situatedness (an experience is often bound to a time and a place), will (experience offers the opportunity for action), and affect (experience is often colored by some mood), among others. However, experiences lacking each of these candidate properties are conceivable---that is, conceiving of them does not lead to self-contradiction or absurdity. They are also achievable, as revealed by altered states of consciousness reached through dreaming, meditative practices, or drugs.

\subsection*{Postulates of physical existence}

To account for the many regularities of experience (Box 1), it is a good inference to assume the existence of a world that persists independently of one’s experience (\textit{realism}). From within consciousness, we can probe the physical existence of things outside of our experience operationally---through observations and manipulations. To be granted physical existence, something should have the power to ``take a difference'' (be affected) and ``make a difference'' (produce effects) in a reliable way (\textit{physicalism}). IIT also assumes operational reductionism: ideally, to establish what exists in physical terms, one would start from the smallest units that can take and make a difference, so that nothing is left out (\textit{atomism}). 

By characterizing physical existence operationally as cause--effect power, IIT can proceed to translate the axioms of phenomenal existence into postulates of physical existence. This establishes the requirements for the \textit{substrate of consciousness}, where ``substrate" is meant operationally as a set of units that can be observed and manipulated. 

\begin{description}
\item[Existence] The substrate of consciousness must have \textit{cause--effect power}: its units must \textit{take and make a difference}.
\end{description}

Building from this ``zeroth'' postulate, IIT formulates the five axioms in terms of postulates of physical existence:

\begin{description}
\item[Intrinsicality] The substrate of consciousness must have \textit{intrinsic} cause--effect power: it must take and make a difference \textit{within itself}.

\item[Information] The substrate of consciousness must have \textit{specific} cause--effect power: it must select a specific \textit{cause--effect state}. 

This state is the one with maximal \textit{intrinsic information} ($\ii$), a measure of the difference a system takes or makes over itself for a given cause state and effect state. 

\item[Integration] 
The substrate of consciousness must have \textit{unitary} cause--effect power: it must specify its cause--effect state as \textit{a whole} set of units, \textit{irreducible} to separate subsets of units. 

Irreducibility is measured by \textit{integrated information} ($\varphi$) over the substrate's minimum partition.

\item[Exclusion] 
The substrate of consciousness must have \textit{definite} cause--effect power: it must specify its cause--effect state as \textit{this} set of units. 

This is the set of units that is maximally irreducible, as measured by maximum $\varphi$ ($\varphi^*$). This set is called a \textit{maximal substrate}, also known as \textit{complex} \cite{Oizumi2014, Marshall2022}.

\item[Composition] 
The substrate of consciousness must have \textit{structured} cause--effect power: subsets of its units must specify cause--effect states over subsets of units (\textit{distinctions}) that can overlap with one another (\textit{relations}), yielding a \textit{cause–effect structure} or $\varPhi$-\textit{structure} (“Phi-structure”).

\end{description}

Distinctions and relations, in turn, must also satisfy the postulates of physical existence: they must have cause--effect power, within the substrate of consciousness, in a specific, unitary, and definite way (they do not have components, being components themselves). They thus have an associated $\varphi$ value. The $\varPhi$-structure unfolded from a complex corresponds to the quality of consciousness. The sum total of the $\varphi$ values of the distinctions and relations that compose the $\varPhi$-structure measures its \textit{structured information} $\varPhi$ (“big Phi”) and corresponds to the quantity of consciousness. 

According to IIT, the physical properties characterized by the postulates are necessary and sufficient for an entity to be conscious. They are necessary because they are needed to account for the properties of experience that are essential, in the sense that it is inconceivable for an experience to lack any one of them. They are also sufficient because no additional property of experience is essential, in the sense that it is conceivable for an experience to lack that property. Thus, no additional physical property is a necessary requirement for being a substrate of consciousness. 

The postulates of IIT have been and are being applied to account for the location of the substrate of consciousness in the brain \cite{Tononi2016} and for its loss and recovery in physiological and pathological conditions \cite{Massimini2005, Casarotto2016}.

\subsection*{The explanatory identity between experiences and $\varPhi$-structures}

Having determined the necessary and sufficient conditions for a substrate to support consciousness, IIT proposes an explanatory identity: every property of an experience is accounted for in full by the physical properties of the $\varPhi$-structure unfolded from a maximal substrate (a complex) in its current state, with no further or ``ad hoc'' ingredients. 
That is, there must be a one-to-one correspondence between the way the experience feels and the way distinctions and relations are structured. Importantly, the identity is not meant as a correspondence between the properties of two separate things. Instead, the identity should be understood in an explanatory sense: the intrinsic (subjective) feeling of the experience can be explained extrinsically (objectively, \emph{i.e.}, operationally or physically) in terms of cause--effect power
\bibnote{Strictly speaking, distinctions and relations that can be singled out phenomenally, such as a spot, a book, and so on, correspond, in physical terms, to bundles of distinctions and relations (compound distinctions and relations)---that is, to sub-structures of a $\varPhi$-structure ($\varPhi$-folds) \cite{Haun2019, Ellia2021}. This can be understood in neural terms because attentional mechanisms can only highlight subsets of units, and thereby all the associated distinctions and relations, rather than individual distinctions and relations. In other words, introspection is the starting point for an explanation of experience in physical terms, but it can only go so far. A full explanation can only be provided through a back-and-forth between the properties of a substrate, which can be explored in great detail, and the  properties of experience, which can only be characterized crudely through introspection}.

The explanatory identity has been applied to account for how space feels (spatial extendedness) and which neural substrates may account for it\cite{Haun2019}. Ongoing work is applying the identity to provide a basic account of the feeling of temporal flow \cite{Comolatti_Time} and that of objects \cite{Grasso_Objects}.
\newline

\begin{tcolorbox}[colback=blue!5!white, colframe=white]
  \subsection*{Box 1. Methodological guidelines of IIT}
  \label{box1}
  \subsubsection*{Inference to a good explanation}
  We should generally assume that an explanation is good if it can account for a broad set of facts (\textit{scope}), does so in a unified manner (\textit{synthesis}), can explain facts precisely (\textit{specificity}), is internally coherent (\textit{self-consistency}), is coherent with our overall understanding of things (\textit{system consistency}), is simpler than alternatives (\textit{simplicity}), and can make testable predictions (\textit{scientific validation}). 
  For example, IIT 4.0 aims at expressing the postulates of intrinsicality, information, integration, and exclusion in a self-consistent manner when applied to systems, causal distinctions, and relations (see formulas).
  
  \subsubsection*{Realism}
  We should assume that something exists (and persists) independently of our own experience. This is a much better hypothesis than solipsism, which explains nothing and predicts nothing. Although IIT starts from our own phenomenology, it aims to account for the many regularities of experience in a way that is fully consistent with realism.
  
  \subsubsection*{Operational physicalism}
  To assess what exists independently of our own experience, we should employ an operational criterion: we should systematically observe and manipulate a substrate's units and determine that they can indeed take and make a difference in a way that is reliable and persisting. Doing so demonstrates a substrate's cause--effect power---the signature of physical existence. Ideally, cause--effect power is fully captured by a substrate's transition probability matrix (TPM) \eqref{eq:tpmU}. This assumption is embedded in IIT's zeroth postulate.
  
  \subsubsection*{Operational reductionism (``atomism'')}
  Ideally, we should account for what exists physically in terms of the smallest units we can observe and manipulate, as captured by unit TPMs. Doing so would leave nothing unaccounted for. IIT assumes that in principle it should be possible to account for everything purely in terms of cause--effect power---cause--effect power ``all the way down'' to conditional probabilities between atomic units. Eventually, this would leave neither room nor need to assume intrinsic properties or laws.
  \end{tcolorbox}
  \begin{tcolorbox}[colback=blue!5!white, colframe=white]
  \subsubsection*{Intrinsic perspective}
  When accounting for experience itself in physical terms, existence should be evaluated from the intrinsic perspective of an entity---what exists for the entity itself, not from the perspective of an external observer. 
  This assumption is embedded in IIT’s postulate of intrinsicality and has several consequences. 
  One is that, from the intrinsic perspective, the quality and quantity of existence must be observer-independent and cannot be arbitrary. 
  For instance, information in IIT must be relative to the specific state the entity is in, rather than an average of states as assessed by an external observer. Similarly, it should be evaluated based on the uniform distribution of possible states, as captured by the entity's TPM \eqref{eq:tpmU}, rather than on an observed probability distribution. 
  By the same token, units outside the entity should be treated as fixed background conditions that do not contribute directly to what the system is. 
  The intrinsic perspective also imposes a tension between expansion and dilution (see below and \cite{Barbosa2020, Barbosa2021}): 
  from the intrinsic perspective of a system (or a mechanism within the system), having more units may increase its informativeness (cause--effect power measured as deviation from chance), while at the same time diluting its selectivity (ability to concentrate cause--effect power over a specific state).
\end{tcolorbox} 

\section*{Overview of IIT's framework}

IIT 4.0 aims at providing a formal framework to characterize the cause--effect structure of a substrate in a given state by expressing IIT's postulates in mathematical terms. 
In line with operational physicalism (Box 1), we characterize a substrate by the transition probability function of its constituting units. 

On this basis, the IIT formalism first identifies sets of units that fulfill all required properties of a substrate of consciousness according to the postulates of physical existence.  
First, for a candidate system, we determine a maximal cause--effect state based on the intrinsic information ($\ii$) that the system in its current state specifies over its possible cause states and effect states. 
We then determine the maximal substrate based on the integrated information ($\varphi_s$) of the maximal cause--effect state. 
To qualify as a substrate of consciousness, a candidate system must specify a maximum of integrated information ($\varphi^*_s$) compared to all competing candidate systems with overlapping units.

The second part of the IIT formalism \textit{unfolds} the cause--effect structure specified by a maximal substrate in its current state, its $\varPhi$-\textit{structure}. To that end, we determine the distinctions and relations specified by the substrate's subsets according to the postulates of physical existence. 
Distinctions are cause--effect states specified over subsets of substrate units (\textit{purviews}) by subsets of substrate units (\textit{mechanisms}). Relations are congruent overlaps among distinctions' cause and/or effect states. Distinctions and relations are also characterized by their integrated information ($\varphi_{d}$, $\varphi_{r}$). The $\varPhi$-structure they compose corresponds to the quality of the experience specified by the substrate; the sum of their $\varphi_{d/r}$ values corresponds to its quantity ($\varPhi$). 

While IIT must still be considered as work in progress, having undergone successive refinements, IIT 4.0 is the first formulation of IIT that strives to characterize $\varPhi$-structures completely and to do so based on measures that satisfy the postulates uniquely. For a comparison of the updated framework with IIT 1.0, 2.0, and 3.0, see \ref{A:Comparison}.

\subsection*{Substrates, transition probabilities, and cause--effect power}

IIT takes physical existence as synonymous with having cause--effect power, the ability to take and make a difference. Consequently, a substrate $U$ with state space $\Omega_U$ is operationally defined by its potential interactions, assessed in terms of conditional probabilities (physicalism, Box 1). We denote the complete transition probability function of a substrate $U$ over a system update $u \rightarrow \up$ as 
\begin{equation}
\label{eq:tpmU}
   \T_U \equiv p(\up\mid \uc), \quad \uc, \up \in \Omega_U.
\end{equation}
A substrate in IIT can be described as a stochastic system $U = \{U_1, U_2, \ldots, U_n\}$ of $n$ interacting units with state space $\Omega_U = \prod_i \Omega_{U_i}$ and current state $u \in \Omega_U$.
We assume that the system updates in discrete steps, that the state space $\Omega_U$ is finite, and that the individual random variables $U_i \in U$ are conditionally independent from each other given the preceding state of $U$: 
\begin{equation}
\label{eq:condind}
p(\up\mid \uc) = \prod_{i=1}^n p(\bar{u}_{i} \mid \uc).
\end{equation}
Finally, we assume a complete description of the substrate, which means that we can determine the conditional probabilities in \eqref{eq:condind} for every system state,
with $p(\up \mid \uc) = p(\up\mid \operatorname{do}(\uc))$ \cite{Albantakis2019, Janzing2013, Ay2008, Pearl2000}, where the ``do-operator'' $\operatorname{do}(\uc)$ indicates that $\uc$ is imposed by intervention. 
This implies that $U$ must correspond to a causal network \cite{Albantakis2019}, and $\T_U$ is a transition probability matrix (TPM) of size $|\Omega_U|$ \bibnote{While the IIT formalism can be applied to hypothetical or simulated systems (as we do for the example systems in the Results/Discussion section), for the resulting quantities to capture existence in physical terms they must be applied to substrate units that can actually be observed and manipulated in physical terms}.

The TPM $\T_U$, which forms the starting point of IIT's analysis, serves as an overall description of a system's cause--effect power: what is the probability that the system will transition into each of its possible states upon an intervention that initializes it into every possible state (Figure \ref{fig:SystemPhi})?  
(Notably, there is no additional role for intrinsic physical properties or laws of nature.) In practice, a causal model will be neither complete nor atomic (capturing the smallest units that can be observed and manipulated), but will capture the relevant features of what we are trying to explain and predict \bibnote{As demonstrated in \cite{AlbantakisQIIT}, it is possible to extend IIT's causal framework to finite quantum systems under unitary evolution, where the conditional independence assumption \eqref{eq:condind} applies to non-entangled subsystems}.

In the ``Results and discussion'' section, the IIT formalism will be applied to extremely simple, simulated networks, rather than causal models of actual substrates. The cause--effect structures derived from these simple networks only serve as convenient illustrations of how a hypothetical substrate’s cause--effect power can be unfolded.

\subsection*{Implementing the postulates}
In what follows, our goal is to evaluate whether a hypothetical substrate (also called ``system'') satisfies all the postulates of IIT. To that end, we must verify whether the system has cause--effect power that is intrinsic, specific, integrated, definite, and structured.

\subsubsection*{Existence}
According to IIT, existence understood as cause--effect power requires the capacity to both take \textit{and} make a difference (see Box 2, Principle of being). 
On the basis of a complete description of the system in terms of interventional conditional probabilities ($\T_U$) \eqref{eq:tpmU}, cause--effect power can be measured as causal \textit{informativeness}. Cause informativeness measures how much a potential cause increases the probability of the current state, and effect informativeness how much the current state increases the probability of a potential effect (as compared to chance). 

\subsubsection*{Intrinsicality}
Building upon the existence postulate, the intrinsicality postulate further requires that a system exerts cause--effect power \textit{within itself}.  
In general, the systems we want to evaluate are open systems $S \subseteq U$ that are part of a larger ``universe'' $U$. 
From the intrinsic perspective of a system $S$ (see Box 1), the set of the remaining units $W = U \setminus S$ merely act as background conditions, whose state can be considered as fixed. This is enforced by \textit{causally conditioning} the larger TPM ($\T_U$) on the current state $W = w$, which makes $W$ causally inert. 

\subsubsection*{Information}
The information postulate requires that a system's cause--effect power be specific: 
the system in its current state must select a specific cause--effect state for its units. 
Based on the \textit{principle of maximal existence} (Box 2), this is the state for which intrinsic information is maximal---the \textit{maximal cause--effect state}. 
\textit{Intrinsic information} ($\ii$) measures the difference a system takes or makes over itself for a given cause and effect state as the product of informativeness and selectivity. 
As we have seen (existence), \textit{informativeness} quantifies the causal power of a system in its current state as a reduction of uncertainty with respect to chance. 
\textit{Selectivity} measures how much cause--effect power is concentrated over that specific cause or effect state. Selectivity is reduced by uncertainty in the cause or effect state with respect to other potential cause and effect states. 

From the intrinsic perspective of the system, the product of informativeness and selectivity leads to a tension between \textit{expansion} and \textit{dilution}, whereby a system comprising more units may show increased deviation from chance but decreased concentration of cause--effect power over a specific state \cite{Barbosa2020, Barbosa2021}.  

\subsubsection*{Integration}
By the integration postulate, it is not sufficient for a system to have cause--effect power within itself and select a specific cause--effect state: it must also specify its maximal cause--effect state in a way that is irreducible. 
This can be assessed by \textit{partitioning} the set of units that constitute the system into separate parts. The system integrated information ($\varphi_s$) then quantifies how much the intrinsic information specified by the maximal state is reduced due to the partition \bibnote{Note that this notion of irreducibility based on set-partitions differs from typical information-theoretic notions such as redundancy or synergy \cite{Albantakis2019cc, Beer2015, Mediano2019_phiid}}. 
Integrated information is evaluated over the partition that makes the least difference, the \textit{minimum partition} (MIP), in accordance with the \textit{principle of minimal existence} (see Box 2).

Integrated information is highly sensitive to the presence of \textit{fault lines}---partitions that separate parts of a system that interact weakly or directionally \cite{Marshall2022}.

\subsubsection*{Exclusion}
Many overlapping sets of units may have a positive value of integrated information ($\varphi_s$). However, the exclusion postulate requires that the substrate of consciousness must be constituted of a definite set of units, neither less nor more. 
Moreover, units, updates, and states must have a definite grain. 
Operationally, the exclusion postulate is enforced by selecting the set of units that maximizes integrated information over itself ($\varphi_s^*$), based again on the principle of maximal existence (see Box 2). 
That set of units is called a \textit{maximal substrate}, or \textit{complex}.
Over a universal substrate, sets of units for which integrated information is maximal compared to all competing candidate systems with overlapping units can be assessed recursively (by identifying the first complex, then the second complex, and so on).

\subsubsection*{Composition}
Once a complex has been identified, composition requires that we characterize its \textit{cause--effect structure} by considering all its subsets and fully \textit{unfolding} its cause--effect power.  

Usually, causal models are conceived in holistic terms, as state transitions of the system as a whole \eqref{eq:tpmU}, or in reductionist terms, as a description of the individual units of the system and their interactions \eqref{eq:condind} \cite{Albantakis2019cc}. However, to account for the structure of experience, considering only the cause--effect power of the individual units or of the system as a whole would be insufficient \cite{Albantakis2019cc, Grasso2021}. Instead, by the composition postulate, we have to evaluate the system’s cause--effect structure by considering the cause--effect power of its subsets as well as their causal relations.

To contribute to the cause--effect structure of a complex, a system subset must both take \textit{and} make a difference (as required by existence) \textit{within} the system (as required by intrinsicality). 
A subset $M \subseteq S$ in state $M = m$ is called a \textit{mechanism} if it \textit{links} a cause and effect state over subsets of units $Z_{c/e} \subseteq S$, called \textit{purviews}. A mechanism together with the cause and effect state it specifies is called a \textit{causal distinction}.
Distinctions are evaluated based on whether they satisfy all the postulates of IIT (except for composition). For every mechanism, the cause--effect state is the one having maximal intrinsic information ($\ii$), and the cause and effect purviews are those yielding the maximum value of integrated information ($\varphi_d$) within the complex---that is, those that are maximally irreducible. 

Distinctions whose cause or effect states overlap congruently within the system (over the same subset of units in the same state) are \textit{bound} together by \textit{causal relations}. Relations also have an associated value of integrated information ($\varphi_r$), corresponding to their irreducibility. 

The distinctions (and associated relations) that exist for the complex are only those whose cause--effect state is congruent with the cause--effect state of the complex as a whole. Together, those distinctions and relations compose the \textit{cause--effect structure} of the complex in its current state. The cause--effect structure specified by a complex is called a $\varPhi$-\textit{structure}. 
The sum of its distinction and relation integrated information amounts to the structured information ($\varPhi$) of the complex.
\newline

In the following, we will provide a formal account of the IIT analysis. The first part demonstrates how to identify complexes. This requires that we (a) determine the cause--effect state of a system in its current state, (b) evaluate the system integrated information ($\varphi_s$) over that cause--effect state, and (c) search iteratively for maxima of integrated information ($\varphi_s^*$) within a universe. 
The second part describes how the postulates of IIT are applied to unfold the cause--effect structure of a complex. This requires that we identify the causal distinctions specified by subsets of units within the complex and the causal relations determined by the way distinctions overlap, yielding the system's cause--effect structure and its structured information $\varPhi$.

\begin{tcolorbox}[colback=blue!5!white, colframe=white]
  \section*{Box 2. Ontological principles of IIT}
  \label{box2}
  \subsubsection*{Principle of being}
  The \textit{principle of being} states that \textit{to be is to have cause--effect power}. In other words, in physical, operational terms, to exist requires being able to take and make a difference. The principle is closely related to the so-called Eleatic principle, as found in Plato's Sophist dialogue \cite{Plato1997}: ``I say that everything possessing any kind of power, either to do anything to something else, or to be affected to the smallest extent by the slightest cause, even on a single occasion, has real existence: for I claim that entities are nothing else but power.'' A similar principle can be found in the work of the Buddhist philosopher Dharmakīrti: ``Whatever has causal powers, that really exists.'' \cite{sep-dharmakiirti} Note that the Eleatic principle is enunciated as a disjunction (either to do something... \textit{or} to be affected...), whereas IIT's principle of being is presented as a conjunction (take \textit{and} make a difference).
  
  \subsubsection*{Principle of maximal existence}
  The \textit{principle of maximal existence} states that, with respect to an essential requirement for existence, \textit{what exists is what exists the most}. 
  The principle is offered by IIT as a good explanation for why the system state specified by the complex and the cause--effect states specified by its mechanisms are what they are. 
  It also provides a criterion for determining the set of units constituting a complex---the one with maximally irreducible cause--effect power, for determining the subsets of units constituting the distinctions and relations that compose its cause--effect structure, and for determining the units' grain. 
  To exemplify, consider a set of candidate complexes overlapping over the same substrate. By the postulates of integration and exclusion, a complex must be both unitary and definite. By the maximal existence principle, the complex should be the one that lays the greatest claim to existence as \textit{one} entity, as measured by system integrated information ($\varphi_s$). For the same reason, candidate complexes that overlap over the same substrate but have a lower value of $\varphi_s$ are excluded from existence. In other words, if having maximal $\varphi_s$ is the reason for assigning existence as a unitary complex to a set of units, it is also the reason to exclude from existence any overlapping set not having maximal $\varphi_s$.
  
  \subsubsection*{Principle of minimal existence}
  Another key principle of IIT is the \textit{principle of minimal existence}, which complements that of maximal existence. The principle states that, with respect to an essential requirement for existence, \textit{nothing exists more than the least it exists}.
  The principle is offered by IIT as a good explanation for why, given that a system can only exist as one system if it is irreducible, its degree of irreducibility should be assessed over the partition across which it is least irreducible (the minimum partition). 
  Similarly, a distinction within a system can only exist as one distinction to the extent that it is irreducible, and its degree of irreducibility should be assessed over the partition across which it is least irreducible. 
  Moreover, a set of units can only exist as a system, or as a distinction within the system, if it specifies both an irreducible cause and an irreducible effect, so its degree of irreducibility should be the minimum between the irreducibility on the cause side and on the effect side
  \bibnote{A principle of IIT not discussed here is the Principle of becoming, which states that powers become what powers do. That is, conditional probabilities in the TPM update depending on what happens. The principle and some of its implications---examined in \cite{Tononi_OnBeing} and forthcoming work}.
\end{tcolorbox}

\section*{Identifying substrates of consciousness through existence, intrinsicality, information, integration, and exclusion}

Our starting point is a substrate $U$ in current state $U = u$ with TPM $\T_U$ \eqref{eq:tpmU}.
We consider any subset $s \subseteq u$ as a possible complex and refer to a set of units $S \subseteq U$ as a candidate system. (Note that $s$, in a slight abuse of notation, refers to both a subset of units in a state, as well as the state itself.)

By the intrinsicality postulate, the units $W = U \setminus S$ are fixed in their current state $w \in \Omega_W$ throughout the analysis of the candidate system $S$ (\textit{causal conditioning}).
Accordingly, we obtain the TPM $\T_S$ of a candidate system $S$ from its intrinsic transition probability function
\begin{equation}
\label{eq:tpm}
\T_S \equiv p(\sp\mid \sc) = p(\sp\mid \sc, w), \quad \sc, \sp \in \Omega_S, 
\end{equation}
where $w = u \setminus s$.

The intrinsic information $\ii_{c/e}$ is a measure of the intrinsic cause/effect power exerted by a system $S$ in its current state $\sc$ over itself by selecting a specific cause/effect state $\bar{s}$. 
The cause--effect state for which intrinsic information ($\ii_{c}$ and $\ii_{e}$) is maximal is called the maximal cause--effect state $s' = \{s'_c, s'_e\}$.
The integrated information $\varphi_s$ is a measure of the irreducibility of a cause--effect state, compared to the directional system partition $\theta'$ that affects the maximal cause--effect state the least (minimum partition, or MIP).
Systems for which integrated information is maximal ($\varphi_s^*$) compared to any competing candidate system with overlapping units are called maximal substrates, or complexes.

\begin{figure}[h!]
    \includegraphics[width=\columnwidth]{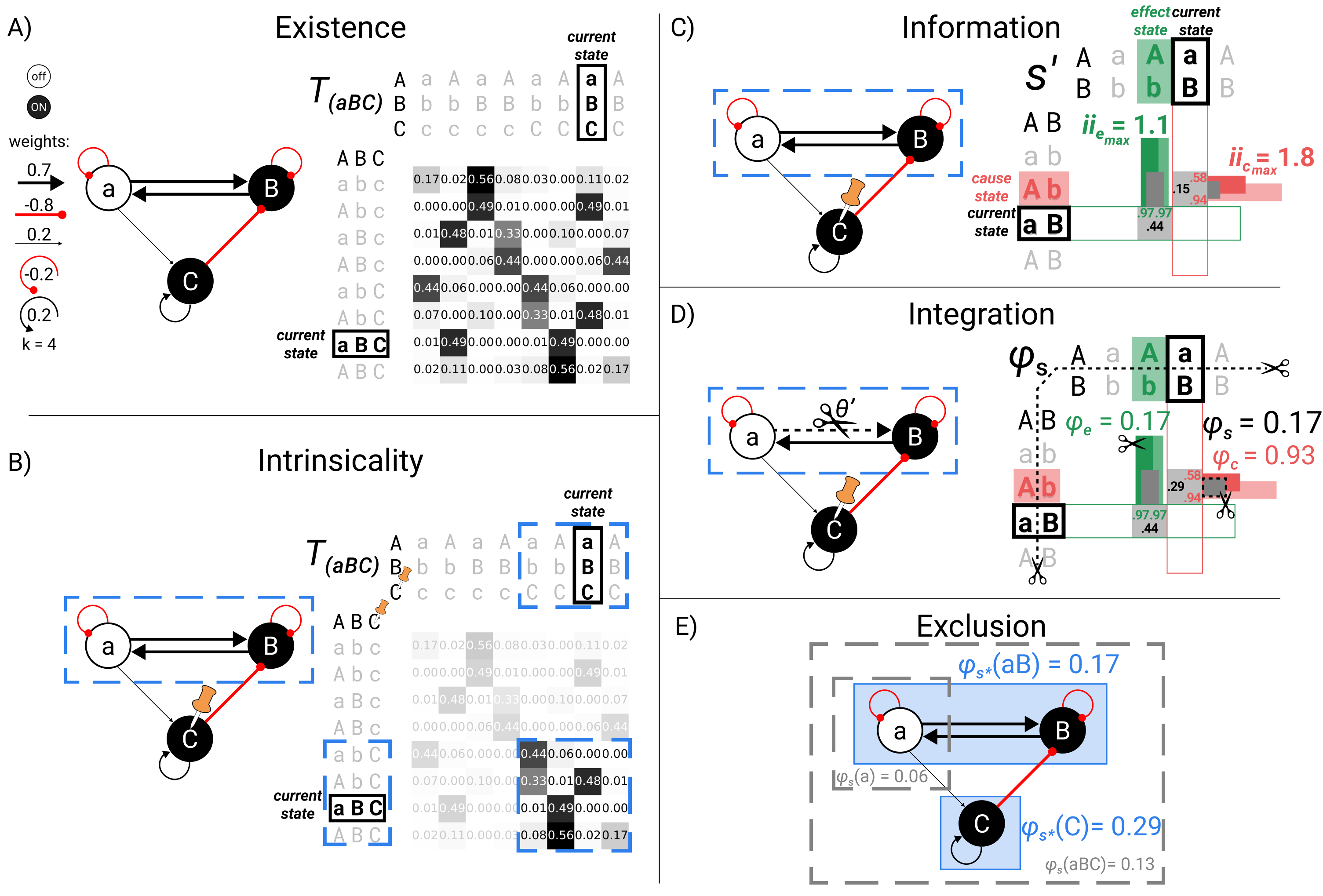}
    \caption{Identifying substrates of consciousness through the postulates of existence, intrinsicality, information, integration, and exclusion.(A) The substrate $S = aBC$ in state $(-1,1,1)$ (lowercase letters for units indicated state ``$-1$'', uppercase letters state ``$+1$'') is the starting point for applying the postulates. The substrate updates its state according to the depicted transition probability matrix (TPM) (each unit follows a logistic equation (see Results for definition) with k = 4.0 and connection weights as indicated in the causal model). Existence requires that the substrate must have cause--effect power, meaning that the TPM among substrate states must differ from chance. (B) Intrinsicality requires that a candidate substrate, for example, units $aB$, has cause--effect power over itself. Units outside the candidate substrate (in this case, unit $C$) are treated as background conditions by pinning them in their current state. (C) Information requires that the candidate substrate $aB$ selects a specific cause--effect state ($s'$). This is the cause state and effect state for which intrinsic information (\emph{$\ii$}) is maximal. Dark-colored and gray bars represent the quantities for informativeness (constrained and unconstrained), and light colored bars for selectivity. (D) Integration requires that the substrate specifies its cause--effect state irreducibly (``as one''). This is established by identifying the minimum partition (MIP) and measuring the integrated information of the system ($\varphi_s$)---the minimum between cause integrated information ($\varphi_c$) and effect integrated information ($\varphi_e$). Here, gray bars represent the partitioned probability. (E) Exclusion requires that the substrate of consciousness is definite, including some units and excluding others. This is established by identifying the candidate substrate with the maximum value of system integrated information ($\varphi^*_s$)---the maximal substrate, or complex. In this case, $aB$ is a complex since its system integrated information ($\varphi_s = 0.17$) is higher than the one of all other overlapping systems (for example, subset $a$ with $\varphi_s = 0.06$ and superset $aBC$ with $\varphi_s = 0.13$).
    }
    \label{fig:SystemPhi}
\end{figure}

The IIT 4.0 formalism to measure a system's integrated information $\varphi_s$ and to identify maximal substrates was first presented in \cite{Marshall2022}. An example of how to identify complexes in a simple system is given in Fig. \ref{fig:SystemPhi}, while a comparison with prior accounts (IIT 1.0, IIT 2.0, and IIT 3.0) can be found in \ref{A:Comparison}.
An outline of the IIT algorithm is included in \ref{A:Algorithm}. 

\subsection*{Existence, intrinsicality, and information: Determining the maximal cause--effect state of a candidate system}

Given a causal model $\T_S$ \eqref{eq:tpm}, we wish to identify the maximal cause--effect state specified by a system in its current state over itself and to quantify the causal power with which it does so.
In doing so, we quantify the cause--effect power of a system from its intrinsic perspective, rather than from the perspective of an outside observer (see Box 1). 

\subsubsection*{System intrinsic information $\ii$}

Intrinsic information $\ii(s,\sp)$ measures the causal power of a system $S$ over itself, for its current state $s$, over a specific cause/effect state $\bar{s}$. Intrinsic information depends on interventional conditional probabilities and unconstrained probabilities of cause/effect states and is the product of selectivity and informativeness. 

On the effect side, intrinsic effect information $\ii_e$ of the current state $\sc$ over a possible effect state $\sp$ is defined as:

\begin{equation}
\label{eq:iise}
     \ii_{\ce}(\sc, \sp) = p_e(\sp \mid \sc) \log\left(\frac{p_e(\sp \mid \sc)}{p_e(\sp)}\right).
\end{equation}

Above, $p_e(\sp \mid \sc) = p(\sp \mid \sc)$ \eqref{eq:tpm} is the interventional conditional probability that the current state $S = \sc$ produces the effect state $\sp$, as indicated by $\T_S$. 

The interventional unconstrained probability $p_e(\sp)$ 

\begin{equation}
\label{eq:pe}
    p_e(\sp) = |\Omega_{S}|^{-1}\sum_{s \in \Omega_{S}} p(\sp \mid \sc),
\end{equation}
is defined as the marginal probability of $\sp$, averaged across all possible current states of $S$ with equal probability (where $|\Omega_{S}|$ denotes the cardinality of the state space $\Omega_{S}$).

On the cause side, intrinsic cause information $\ii_c$ of the current state $\sc$ over a possible cause state $\sm$ is defined as:

\begin{equation}
\label{eq:iisc}
     \ii_{c}(\sc, \sp) = p_c(\sm \mid \sc) \log\left(\frac{p_e(\sc \mid \sm)}{p_e(\sc)}\right).
\end{equation}
Above, $p_c(\sm \mid \sc)$ is the interventional conditional probability that the current state $S = s$ was produced by $\sm$. The latter is derived from $\T_S$ using Bayes' rule, where we again assign a uniform prior to the states of $\sm$,

\begin{equation}
\label{unconstrained_cause}
    p_c(\sm) = |\Omega_{S}|^{-1}, ~ \sm \in \Omega_S,
\end{equation}
such that

\begin{equation}
\label{eq:pc}
    p_c(\sm \mid \sc) =  \frac{p_e(\sc \mid \sm)\cdot |\Omega_{S}|^{-1}}{p_e(\sc)}  = \frac{p(s \mid \bar{s})}{\sum\limits_{\hat{s} \in \Omega_S} p(s \mid \hat{s})}.
\end{equation}

\subsubsection*{Informativeness (over chance)}

In \eqref{eq:iise} and \eqref{eq:iisc}, the logarithmic term (in base 2 throughout) is called \textit{informativeness}. 
Note that informativeness is expressed in terms of effect probabilities for both $\ii_e$ \eqref{eq:iise} and $\ii_c$ \eqref{eq:iisc}. However, $\ii_e$ \eqref{eq:iise} evaluates the increase in probability of the effect state due to the current state, while $\ii_c$ \eqref{eq:iisc} evaluates the increase in probability of the current state due to the cause state.

In line with the existence postulate, a system $S$ in state $\sc$ has cause--effect power (it takes and makes a difference) if it raises the probability of a possible effect state compared to chance, which is to say compared to its unconstrained probability, 

\begin{equation}
    \log\left(\frac{p_{e}(\bar{s} \mid \sc)}{p_{e}(\bar{s})}\right) > 0,
\end{equation}
and if the probability of the current state is raised above chance by a possible cause state,
\begin{equation}
    \log\left(\frac{p_{e}(\sc \mid \sm)}{p_{e}(\sc)}\right) > 0.
\end{equation}
Informativeness is additive over the number of units: if a system specifies a cause or effect state with probability $p=1$, its causal power increases additively with the number of units whose states it fully specifies (\textit{expansion}), given that the chance probability of all states decreases exponentially.

\subsubsection*{Selectivity (over states)}

From the intrinsic perspective of a system, cause--effect power over a specific cause or effect state depends not only on the deviation from chance it produces, but also on how that deviation is concentrated on that state, rather than being diluted over other states. 
This is measured by the \textit{selectivity} term in front of the logarithmic term in \eqref{eq:iise} and \eqref{eq:iisc}, corresponding to the conditional probability $p_{c/e}(\bar{s}∣\sc)$ of that specific cause or effect state. 
Selectivity means that if $p<1$, the system’s causal power becomes subadditive (\textit{dilution}) (see \cite{Barbosa2020} for details).
For example, as shown in \cite{Barbosa2021}, if an unconstrained unit is added to a fully specified unit, intrinsic information does not just stay the same, but decreases exponentially. 
From the intrinsic perspective of the system, the informativeness of a specific cause/effect state is diluted because it is spread over multiple possible states, yet the system must select only one state.

Altogether, taking the product of informativeness and selectivity leads to a tension between expansion and dilution: a larger system will tend to have higher informativeness than a smaller system because it will deviate more from chance, but it will also tend to have lower selectivity because it will have a larger repertoire of states to select from.

Because of the selectivity term, intrinsic information is reduced by indeterminism and degeneracy. As shown in \cite{Marshall2022}, indeterminism decreases the probability of the selected effect state because it implies that the same state can lead to multiple states. In turn, degeneracy decreases the probability of the selected cause state because it implies that multiple states can lead to the same state, even in a deterministic system.

The intrinsic information $\ii$ is quantified in units of \emph{intrinsic bits}, or \emph{ibits}, to distinguish it from standard information-theoretic measures (which are typically additive). Formally, the \emph{ibit} corresponds to a point-wise information value (measured in bits) weighted by a probability. 

\subsubsection*{The maximal cause--effect state}
Taking the product of informativeness and selectivity on the system's cause and effect sides captures the postulates of existence (taking and making a difference) and intrinsicality (taking and making a difference over itself) for each possible cause/effect state, as measured by intrinsic information. However, the information postulate further requires that the system selects a specific cause/effect state. Which one is determined based on the principle of maximal existence (Box 1): the cause/effect specified by the system should be the one that maximizes intrinsic information. On the effect side (and similarly for the cause side),

\begin{equation}
\label{eq:s'}
    \begin{split}
        s'_{\ce}(\T_S, \sc) =& \argmax_{\sp \in \Omega_S} \ii_{\ce}(\sc, \sp) \\ 
             =& \argmax_{\sp \in \Omega_S} p_e(\sp \mid \sc)\log\left(\frac{p_e(\sp \mid \sc)}{p_e(\sp)}\right).
    \end{split}
\end{equation}
The system's intrinsic effect information is the value of $\ii_e$ \eqref{eq:iise} for its maximal effect state:

\begin{equation}
\label{eq:iis'}
   \ii_{\ce}(\T_S, s) :=\ii_{\ce}(s, s'_e) = \max_{\sp \in \Omega_S} p_e(\sp \mid \sc)\log\left(\frac{p_e(\sp \mid \sc)}{p_e(\sp)}\right).
\end{equation}
We have made the dependency of $s'$ and $\ii_e$ on $\T_S$ explicit in \eqref{eq:s'} and \eqref{eq:iis'} to highlight that for intrinsic information to properly assess cause--effect power, all probabilities must be derived from the system's interventional transition probability function, while imposing a uniform prior distribution over all possible system states. 
If there is no state with $\ii_e(\T_S, s) > 0$, the system $S$ in state $s$ has no causal power (and likewise for $\ii_c(\T_S,s)$).
Note also that a system's intrinsic cause/effect state does not necessarily correspond to the actual cause/effect state (what actually happened before / will happen after) in the dynamical evolution of the system, which typically also depends on extrinsic influences. 
(For an account of actual causation according to the causal principles of IIT, see \cite{Albantakis2019}.) 

Because consciousness is the way it is, the translation of its properties in physical, operational terms should be unique and based on quantities that uniquely satisfy the postulates \cite{Barrett2019, Barbosa2021}.
Intrinsic information \eqref{eq:iis'} is formally equivalent to a measure of intrinsic difference \cite{Barbosa2020}, which uniquely satisfies three desired properties (causality, specificity, and intrinsicality) that align with the postulates of IIT, but also have independent justification. 

\subsection*{Integration: Determining the irreducibility of a candidate system}

Having identified the maximal cause--effect state $s' = \{s'_c, s'_e\}$ of a candidate system $S$ in its current state $s$, the next step is to evaluate whether the system specifies the cause--effect state of its units in a way that is \textit{irreducible}, as required by the integration postulate: a candidate system can only be a substrate of consciousness if it is \textit{one} system---that is, if it cannot be subdivided into subsets of units that exist separately from one another.

\subsubsection*{Directional system partitions}

To that end, we define a set of \textit{directional} system partitions $\Theta(S)$ that divide $S$ into $k \geq 2$ parts $\{S^{(i)}\}^k_{i=1}$, such that 

\begin{equation}
    S^{(i)} \neq \varnothing, \ S^{(i)} \cap S^{(j)} = \varnothing, \mathrm{\ and\ } \bigcup_{i = 1}^k S^{(i)} = S.
\end{equation}
In words, each part $S^{(i)}$ must contain at least one unit, there must be no overlap between any two parts $S^{(i)}$ and $S^{(j)}$, and every unit of the system must appear in exactly one part.
For each part $S^{(i)}$, the partition removes the causal connections of that part with the rest of the system in a directional manner: either the part's inputs, outputs, or both are replaced by independent ``noise'' (they are ``cut'' by the partition in the sense that their causal powers are substituted by chance). 
Directional partitions are necessary because, from the intrinsic perspective of a system, a subset of units that cannot affect the rest of the system, or cannot be affected by it, cannot truly be a part of the system. In other words, to be a part of a system, a subset of units must be able to interact with the rest of the system in both directions (cause \textit{and} effect). 

A partition $\theta \in \Theta(S)$ thus has the form

\begin{equation}
    \theta = \{S^{(1)}_{\delta_1}, S^{(2)}_{\delta_2}, \ldots,  S^{(k)}_{\delta_k}\},
\end{equation}
where $\delta_i \in \{\inp, \out, \both\}$ indicates whether the inputs ($\inp$), outputs ($\out$), or both ($\both$) are cut for a given part. 
For each part $S^{(i)}$, we can then identify a set of units $X^{(i)} \subseteq S$ whose inputs to $S^{(i)}$ have been cut by the partition, and the complementary set $Y^{(i)} = S \setminus X^{(i)}$ whose inputs to $S^{(i)}$ are left intact.
Specifically,

\begin{equation}
    X^{(i)} = \begin{cases} S \setminus S^{(i)} & \text{ if } \delta_i \in \{\inp, \both\} \\ & \\ \bigcup\limits_{\substack{j \neq i: \\ \delta_j \in \{\out, \both\}}} S^{(j)} & \text{ if } \delta_i \in \{\out\}. \end{cases}
\end{equation}

Given a partition $\theta \in \Theta(S)$, we define a partitioned transition probability matrix $\T_S^\theta$ in which all connections affected by the partition are ``noised.'' This is done by combining the independent contributions of each unit $S_j \in S$ in line with the conditional independence assumption \eqref{eq:condind}, such that

\begin{equation}
\label{eq:tpmSpart}
\T_S^\theta \equiv p^\theta(\sp\mid \sc) = \prod_{j=1}^n p^\theta(\bar{s}_{j} \mid \sc), ~ \sp, \sc \in \Omega_S,
\end{equation}
where the partitioned probability of a unit $S_j \in S^{(i)}$ is defined as
\begin{equation}
    p^\theta(\bar{s}_{j} \mid \sc) = |\Omega_{X^{(i)}}|^{-1} \sum_{x \in \Omega_{X^{(i)}}} p(\bar{s}_{j} \mid x, Y^{(i)} = y).
\end{equation}
This means that all connections to unit $S_j$ that are affected by the partition are \textit{causally marginalized} (replaced by independent noise). 

\subsubsection*{System integrated information $\varphi_s$}

The integrated effect information $\varphi_{\ce}$ measures how much the partition $\theta \in \Theta_S$ reduces the probability with which a system $S = s$ specifies its effect state $s'_{\ce}$ \eqref{eq:s'},

\begin{equation}
\label{eq:phise}
    \varphi_{\ce}(\T_S, s,\theta) =  p_e(s'_{\ce} \mid s)\log\left(\frac{p_e(s'_{\ce} \mid s)}{p_e^\theta(s'_{\ce}\mid s)}\right).
\end{equation}
%
Note that $\varphi_{\ce}$ has the same form as the intrinsic information $\ii_{\ce}(s, \sp)$ \eqref{eq:iise}, with the partitioned effect probability taking the place of the unconstrained (marginal) probability.
Likewise, the integrated cause information $\varphi_c$ is defined as 

\begin{equation}
\label{eq:phisc}
    \varphi_{c}(\T_S, s,\theta) =  p_c(s'_{c} \mid s)\log\left(\frac{p_e(s \mid s'_c)}{p_e^\theta(s\mid s'_c)}\right).
\end{equation}
(By the principle of maximal existence, if two or more cause--effect states are tied for maximal intrinsic information, the system specifies the one that maximizes $\varphi_{c/e}$.) 

By the zeroth postulate, existence requires cause \textit{and} effect power, and the integration postulate requires that its cause--effect power be irreducible. By the principle of minimal existence (Box 2), then, system integrated information for a given partition is the minimum of its irreducibility on the cause and effect sides: 

\begin{equation}
\label{eq:phis}
    \varphi_s(\T_S,s,\theta) = \min \{\varphi_c(\T_S,s,\theta), \varphi_e(\T_S,s, \theta)\}.
\end{equation}
Accordingly, the system is reducible if at least one partition $\theta \in \Theta_S$ makes no difference to the cause or effect probability.

Moreover, again by the principle of minimal existence, the integrated information of a system is given by its irreducibility over its minimum partition (MIP) $\theta' \in \Theta_S$, such that

\begin{equation}
\label{eq:phis'}
    \varphi_s(\T_S,s) :=  \varphi_s(\T_S,s, \theta').
\end{equation}

The MIP is defined as the partition $\theta \in \Theta_S$ that minimizes the system's integrated information, relative to the maximum possible value it could take for an arbitrary TPM $\T'_{S}$ over the units of system $S$

\begin{equation}
\label{eq:mips}
    \theta' = \argmin_{\theta \in \Theta(S)} \frac{\varphi_s(\T_S, s, \theta)}{\displaystyle\max_{\T'_S} \varphi(\T'_S,s, \theta)}.
\end{equation}
The normalization ensures that $\varphi_s(\T_S,s)$ is evaluated fairly over a system's fault lines by assessing integration relative to its maximum possible value over a given partition. 
Using the \textit{relative} integrated information quantifies the strength of the interactions between parts in a way that does not depend on the number of parts and their size.
As proven in \cite{Marshall2022}, the maximal value of $\varphi(\T_S,s,\theta)$ for a given partition $\theta$ is the normalization factor $\displaystyle\max_{\T'_S} \varphi(\T_S,s, \theta) = \sum_{i = 1}^k |S^{(i)}||X^{(i)}|$, which corresponds to the maximal possible number of ``connections'' (pairwise interactions) affected by $\theta$.
For example, as shown in \cite{Marshall2022}, the MIP will correctly identify the fault line dividing a system into two large subsets of units linked through a few interconnected units (a ``bridge''), rather than defaulting to partitions between individual units and the rest of the system.
Once the minimum partition has been identified, the integrated information across it is an \textit{absolute} quantity, quantifying the loss of intrinsic information due to cutting the minimum partition of the system. 
(If two or more partitions $\theta \in \Theta(S)$ minimize Eqn. \eqref{eq:mips}, we select the partition with the largest unnormalized $\varphi_s$ value as $\theta'$, applying the principle of maximal existence.)
Defining $\theta'$ as in \eqref{eq:mips}, moreover, ensures that $\varphi_s(\T_S,s) = 0$ if the system is not strongly connected in graph-theoretic terms. 

In summary, the system integrated information ($\varphi_s(\T_S,s)$) quantifies the extent to which system $S$ in state $s$ has cause--effect power over itself as one system (\emph{i.e.}, irreducibly). $\varphi_s(\T_S,s)$ is thus a quantifier of irreducible existence.

\subsection*{Exclusion: Determining maximal substrates (complexes)}

In general, multiple candidate systems with overlapping units may have positive values of $\varphi_s(\T_S,s)$. By the exclusion postulate, the substrate of consciousness must be definite; that is, it must comprise a definite set of units. But which one? 
Once again, we employ the principle of maximal existence (Box 2): among candidate systems competing over the same substrate with respect to an essential requirement for existence, in this case irreducibility, the one that exists is the one that exists the most. Accordingly, the maximal substrate, or complex, is the candidate substrate with the maximum value of system integrated information ($\varphi^*_s$), and overlapping substrates with lower $\varphi_s$ are excluded from existence.

\subsubsection*{Determining maximal substrates recursively}
Within a universal substrate $U_0$ in state $u_0$, subsets of units that specify maxima of irreducible cause--effect power (complexes) can be identified by an iterative search for the system $S_k^* \subseteq U_k$ with
\begin{equation}
    \varphi_s^*(\T_{U_k},u_k) = \max_{S \subseteq U_k} \varphi_s(\T_S, s),
\end{equation}
such that
\begin{equation}
\label{eq:sk}
    S_k^* = \argmax_{S \subseteq U_k} \varphi_s(\T_S, s),
\end{equation}
and $U_{k+1} = U_k \setminus S_k^*$ until $U_{k+1} = \emptyset$ or $U_{k+1} = u_k$ (the units in $U_0 \setminus U_{k+1}$ still serve as background conditions). If the maximal substrate $S^*_k$ is not unique, and all tied systems overlap, the next best system that is unique is chosen instead; for details see \cite{Marshall2022}, and also \cite{Moon2019}.

For any complex $S^* = s^*$, overlapping substrates that specify less integrated information ($\varphi_s < \varphi_s(\T_{S^*}, s^*)$) are excluded. Consequently, specifying a maximum of integrated information $\varphi_s^*$ compared to all overlapping systems 
\begin{equation}
\label{eq:complex_rec}
    S \cap \Tilde{S} \neq \varnothing \Rightarrow \varphi_s(s) > \varphi_s(\Tilde{s}), \;\; \forall \Tilde{S}\neq S \subseteq U
\end{equation}
is a sufficient requirement for a system $S \subseteq U$ to be a complex.

As shown in \cite{Marshall2022}, this recursive search for maximal substrates ``condenses'' the universe $U_0 = u_0$ into a disjoint (non-overlapping) and exhaustive set of complexes---the first complex, second complex, and so on.

\subsubsection*{Determining maximal unit grains}
Above, we presented how we can determine the borders of a complex within a larger system $U$, assuming a particular spatio-temporal grain for the units $U_i \in U$. 
In principle, however, all possible grains should be considered \cite{Hoel2016, Marshall2018}. 
In the brain, for example, the grain of units could be brain regions, groups of neurons, individual neurons, sub-cellular structures, molecules, atoms, quarks, or anything finer, down to hypothetical atomic units of cause--effect power \cite{Tononi2015_scholarpedia, Tononi2016}. For any unit grain---neurons, for example---the grain of updates could be minutes, seconds, milliseconds, micro-seconds, and so on. And the grain of states could be 2 states (``no spikes / any number of spikes per neuron over one hundred milliseconds''), 4 states (``no spikes, 1 spike, 2--5 spikes, bursts of $>5$ spikes''), or 256 states (from 0 to 255 spikes in intervals of 1 spike), and so on. 
However, by the exclusion postulate, the units that constitute a system $S$ must also be definite, in the sense of having a definite spatio-temporal grain. 

Once again, the grain is defined by the principle of maximal existence: across the possible micro- and macroscopic levels, the ``winning'' grain is the one that ensures maximally irreducible existence ($\varphi_s^*$) for the entity to which the units belong \cite{Hoel2016, Marshall2018}. 

As constituents of a complex upon which its cause--effect power rests, the units themselves should comply with the postulates of IIT \cite{Tononi_OnBeing}. Otherwise it would be possible to ``make something out of nothing.'' Accordingly, units themselves must also be maximally irreducible, as measured by the unit's integrated information ($\varphi_j \equiv \varphi_s$); otherwise, they would not be units but ``disintegrate'' into their constituents. 
However, in contrast to systems, units only need to be maximally irreducible within, because they do not exist as entities in their own right: 
a unit $J$ qualifies as a candidate unit of a larger system $S$ if its integrated information $\varphi(j)$ is higher than that of any of its subsets
\begin{equation}
\label{eq:unit_rec}
    \Tilde{J} \subset J \Rightarrow \varphi_s(\T_J, j) > \varphi_s(\T_{\Tilde{J}}, \Tilde{j}), \;\; \forall \Tilde{J} \subset J.
\end{equation}
Out of all possible sets of such candidate units, the set of (macro) units that actually exists is the one that maximizes the existence of the complex to which they belong, rather than their own existence.

\section*{Unfolding the cause--effect structure of a complex through composition}

Once a maximal substrate and the associated maximal cause--effect state have been identified, we must unfold its cause--effect power to reveal its cause--effect structure of distinctions and relations, in line with the composition postulate. 
As components of the cause--effect structure, distinctions and relations must also satisfy the postulates of IIT (save for composition). 

\subsection*{Composition and causal distinctions}

Causal distinctions capture how the cause--effect power of a substrate is structured by subsets of units that specify irreducible causes and effects over subsets of its units.
A candidate distinction $d(m)$ consists of
\begin{enumerate*}[label=(\arabic*)]
    \item a mechanism $M \subseteq S$ in state $m \in \Omega_M$ inherited from the system state $s \in \Omega_S$; 
    \item a maximal cause--effect state $z^* = \{z^*_c, z^*_e\}$ over the cause and effect purviews ($Z_c, Z_e \subseteq S$) linked by the mechanism; and 
    \item an associated value of irreducibility ($\varphi_d > 0$).
\end{enumerate*}
A distinction $d(m)$ is thus represented by the tuple
\begin{equation}
\label{eq:distinction}
    d(m) = \Big(m, z^*, \varphi_d\Big).
\end{equation}

For a given mechanism $m$, our goal is to identify the maximal cause $Z^*_c = z^*_c$ and effect $Z^*_e = z^*_e$ of $m$ within the system, where $Z^*_c, Z^*_e \subseteq S$, $z^*_c \in \Omega_{Z^*_c}$, $z^*_e \in \Omega_{Z^*_e}$. 

As above, in line with existence, intrinsicality, and information, we determine the maximal cause or effect state specified by the mechanism over a candidate purview within the system based on the value of intrinsic information $\ii(m, z)$.
Next, in line with integration, we determine the value of integrated information $\varphi_d(m, Z, \theta)$ over the minimum partition $\theta'$. 
In line with exclusion, we determine the maximal cause--effect purviews for that mechanism over all possible purviews 
$Z \subseteq S$ based on the associated value of irreducibility $\varphi_d(m, Z, \theta')$.  Finally, we determine whether the maximal cause--effect state specified by the mechanism is congruent with the system's overall cause--effect state ($z^*_c \subseteq s^*_c$, $z^*_e \subseteq s^*_e$), in which case we conclude that it contributes a distinction to the overall cause--effect structure.

The updated formalism to identify causal distinctions within a system $S$ in state $s$ was first presented in \cite{Barbosa2021}. Here we provide a summary with minor adjustments on selecting $z^*_c$ and $z^*_e$, the cause integrated information $\varphi_c(m,Z)$, and the requirement that causal distinctions must be congruent with the system's maximal cause--effect state (see \ref{A:Comparison}). 

\subsubsection*{Existence, intrinsicality, and information: Determining the cause and effect state specified by a mechanism over candidate purviews}

Like the system as a whole, its subsets must comply with existence, intrinsicality, and information. As for the system, we begin by quantifying, in probabilistic terms, the difference a subset of units $M \subseteq S$, in its current state $m \subseteq s$ takes and makes from and to subsets of units $Z\subseteq S$ (cause and effect purview). As above, we start by establishing the interventional conditional probabilities and unconstrained probabilities from the TPM $\T_S$. 

When dealing with a mechanism constituted by a subset of system units, it is important to capture the constraints on a purview state $z$ that are exclusively due to the mechanism in its state ($m$), removing any potential contribution from other system units.   
This is done by causally marginalizing all variables in $X = S\setminus M$, which corresponds to imposing a uniform distribution as $p(X_t)$ \cite{Oizumi2014, Albantakis2019, Barbosa2021} (note \bibnote{Units within the candidate system are causally marginalized to discount their causal contribution if they are not part of the mechanism or purview under consideration. By contrast, units outside the candidate system are causally conditioned in their current state throughout the analysis and act as fixed background conditions}). 
In addition, product probabilities $\pi(z \mid m)$ are used instead of conditional probabilities $p(z\mid m)$ to discount correlations from units in $X = S\setminus M$ with divergent outputs to multiple units in $Z \subseteq S$ \cite{Oizumi2014, Krohn2017, Albantakis2019}.
In this way, causal marginalization maintains the conditional independence between units \eqref{eq:condind} by applying independent noise to individual connections.
The assumption of conditional independence distinguishes IIT's causal analysis from standard information-\allowbreak theoretic analyses of information flow \cite{Ay2008, Albantakis2019} and corresponds to an assumption that variables are ``physical'' units in the sense that they are irreducible within and can be observed and manipulated independently.

The effect probability of a single unit $Z_i \in Z$ conditioned on the current state $m$ is thus defined as

\begin{equation}
\label{eq:cmarg}
\pi_e(z_i \mid m) = p_e(z_i \mid m) = |\Omega_{X}|^{-1} \sum_{x \in \Omega_{X}} p\left(z_{i} \mid m,x\right), \quad z_{i} \in \Omega_{Z_i} 
\end{equation}
and the effect probability over a set $Z$ of $|Z|$ units is defined as the product of the effect probabilities over individual units 

\begin{equation}
\label{eq:effect_probability}
\pi_e(z \mid m) = \prod_{i=1}^{|Z|} \pi_e(z_i \mid m), \quad z \in \Omega_{Z}.
\end{equation}
From Eq. \eqref{eq:effect_probability} we can also define an unconstrained effect probability

\begin{equation}
\label{eq:UCE}
    \pi_e(z;M) = |\Omega_{M}|^{-1} \sum_{m \in \Omega_{M}} \pi_e(z \mid M = m) , \quad z \in \Omega_{Z}.
\end{equation}

Given the set $Y = S \setminus Z$, the cause probability for a mechanism $m$ with $|M|$ units is computed using Bayes' rule over the product distributions

\begin{equation}
\label{eq:cause_probability}
\pi_c(z \mid m) = \frac{\displaystyle
    \pi_e(m \mid z) \cdot |\Omega_{Z}|^{-1}
}{\displaystyle
    \pi_e(m; Z)
} = \frac{\displaystyle
    \prod_{i=1}^{|M|} p_e(m_{i} \mid z) 
}{\displaystyle
    \sum_{\hat{z} \in \Omega_{Z}} \prod_{i=1}^{|M|} p_e(m_{i} \mid \hat{z}) 
}, \quad z \in \Omega_{Z},
\end{equation}
where $\displaystyle p_e(m_{i} \mid z) = |\Omega_{Y}|^{-1}\sum_{y \in \Omega_Y} p(m_{i} \mid z, y)$ in line with \eqref{eq:cmarg}.

To correctly quantify intrinsic causal constraints, the unconstrained cause probability is again set to the uniform distribution $\pi_c(z)= |\Omega_{Z}|^{-1}, \; z \in \Omega_{Z}$ (unlike the unconstrained effect probability, the unconstrained cause distribution does not depend on the mechanism).
Note that the product in \eqref{eq:cause_probability} is over units of $M$, not of $Z$. 
For details see \cite{Albantakis2019, Barbosa2021}. 
As above, all probabilities $p$ are obtained from the TPM $\T_S$ \eqref{eq:tpm} and thus correspond to \textit{interventional} probabilities throughout.

Having defined cause and effect probabilities, we can now evaluate the intrinsic information of a mechanism $m$ over a purview state $Z = z$ analogously to the system intrinsic information \eqref{eq:iise} and \eqref{eq:iisc}. 
The intrinsic effect information that a mechanism in a state $m$ specifies about a purview state $z$ is 
\begin{equation}
\label{eq:iie}
    \ii_e(m, z) = \pi_e(z \mid m)\log\left(\frac{\pi_e(z \mid m)}{\pi_e(z; M)}\right).
\end{equation}
The intrinsic cause information that a mechanism in a state $m$ specifies about a purview state $z$ is
\begin{equation}
\label{eq:iic}
    \ii_c(m, z) = \pi_c(z \mid m)\log\left(\frac{\pi_e(m \mid z)}{\pi_e(m; Z)}\right).
\end{equation}

As with system intrinsic information, the logarithmic term is the informativeness, which captures how much causal power is exerted by the mechanism $m$ on its potential effect $z$ (how much it increases the probability of that state above chance), or by the potential cause $z$ on the mechanism $m$.
The first term corresponds to the mechanism's selectivity, which captures how much the causal power of the mechanism $m$ is concentrated on a specific state of its purview (as opposed to other states). 
In the following we will again focus on the effect side, but an equivalent procedure applies on the cause side.

Based on the principle of maximal existence, the maximal effect state of $m$ within the purview $Z$ is defined as  
\begin{equation}
\label{eq:effect_state}
\begin{split}
    z'_e(m, Z) =& \argmax_{z \in \Omega_Z} \ii_{e}(m,z)\\
    =&\argmax_{z \in \Omega_Z} \left( \pi_e(z \mid m)\log\left(\frac{\pi_e(z \mid m)}{\pi_e(z;M)}\right) \right),
\end{split}
\end{equation}
which corresponds to the specific effect of $m$ on $Z$. Note that $z'_e$ is not always unique (see \ref{A:Ties}). 
The maximal intrinsic information of mechanism $m$ over a purview $Z$ is then 
\begin{equation}
\label{eq:iie'}
    \ii_e(m, Z) := \ii_e(m,z'_e) = \max_{z \in \Omega_Z} \ii_e(m,z). 
\end{equation}

The intrinsic information of a candidate distinction, like that of the system as a whole, is sensitive to indeterminism (the same state leading to multiple states) and degeneracy (multiple states leading to the same state) because both factors decrease the probability of the selected state. Moreover, the product of selectivity and informativeness leads to a tension between expansion and dilution: larger purviews tend to increase informativeness because conditional probabilities will deviate more from chance, but they also tend to decrease selectivity because of the larger repertoire of states.

\subsubsection*{Integration: Determining the irreducibility of a candidate distinction}

To comply with integration, we must next ask whether the specific effect of $m$ on $Z$ is irreducible. As for the system, we do so by evaluating the integrated information $\varphi_e(m,Z)$. To that end, we define a set of ``disintegrating'' partitions $\Part(M,Z)$ as
\begin{adjustwidth}{-0.75in}{0in}
\begin{multline}
\label{eq:partitions}
\Part(M,Z) = \Bigg\{ \{(M^{(i)}, Z^{(i)})\}_{i=1}^{k} \; : \; k \in \{2,3,4,\ldots\},\; M^{(i)} \in \powerset(M),\; Z^{(i)} \in \powerset(Z), 
\\ 
\bigcup M^{(i)} = M, \bigcup Z^{(i)} = Z, 
Z^{(i)} \cap Z^{(j)} = M^{(i)} \cap M^{(j)} = \O \; \forall \; i \neq j, M^{(i)} = M \Longrightarrow Z^{(i)} = \O\Bigg\},
\end{multline}
\end{adjustwidth}
where $\{M^{(i)}\}$ is a partition of $M$ and $\{Z^{(i)}\}$ is a partition of $Z$, but the empty set may also be used as a part ($\powerset$ denotes the power set).
As introduced in \cite{Albantakis2019, Barbosa2021}, a disintegrating partition $\p \in \Part(M,Z)$ either ``cuts'' the mechanism into at least two independent parts if $|M| > 1$, or it severs all connections between $M$ and $Z$, which is always the case if $|M| = 1$ (we refer to \cite{Albantakis2019, Barbosa2021} for details). 

Given a partition $\p \in \Part(M,Z)$, we can define the partitioned effect probability 
\begin{equation}
    \label{eq:partitioned_effect}
    \pi_e^{\theta}(z'_e \mid m) = \prod_{i = 1}^k \pi_e(z'^{(i)}_{e} \mid m^{(i)}),
\end{equation}
with $\pi(\varnothing|m^{(i)}) = \pi(\varnothing) = 1$. In the case of $m^{(i)} = \varnothing$,  $\pi_e(z'^{(i)}_e|\varnothing)$ corresponds to the fully partitioned effect probability

\begin{equation}
\label{eq:FPCE}
    \pi_e(z\mid \varnothing) = \prod_{i=1}^{|Z|} \pi_e(z_{i}) = \prod_{i=1}^{|Z|} \sum_{\sc \in \Omega_{S}} p_e(z_{i} \mid \sc) |\Omega_{S}|^{-1}.
\end{equation}

The integrated effect information of mechanism $m$ over a purview $Z \subseteq S$ with effect state $z'_e$ for a particular partition $\theta \in \Theta(M,Z)$ is then defined as
\begin{equation}
\varphi_e(m,Z, \theta) = \pi_e(z'_e \mid m)\log\left(\frac{\pi_e(z'_e \mid m)}{\pi_e^\theta(z'_e \mid m)}\right).
\end{equation}
The effect of $m$ on $z'_e$ is reducible if at least one partition $\p \in \Part(M,Z)$ makes no difference to the effect probability. 
In line with the principle of minimal existence, the total integrated effect information $\varphi_e(m, Z)$ again has to be evaluated over $\theta'$, the minimum partition (MIP)

\begin{equation}
\label{eq:phieZ}
    \varphi_e(m, Z) := \varphi_e(m, Z, \theta'),
\end{equation}
which requires a search over all possible partitions $\p \in \Part(M,Z)$:

\begin{equation}
\label{eq:mip}
\p' = \argmin_{\p \in \Theta(M,Z)} \frac{\varphi(m,Z,\theta)}{\displaystyle\max_{\T'_S} \varphi(m,Z,\theta)}.
\end{equation}
As in \eqref{eq:mips}, the minimum partition is evaluated against its maximum possible value across all possible system $\T'_S$, which again corresponds to the number of possible pairwise interactions affected by the partition.

The integrated cause information is defined analogously, as 

\begin{equation}
\varphi_c(m,Z) := \varphi_c(m,Z, \theta') = \pi_c(z'_c \mid m)\log\left(\frac{\pi_e(m \mid z'_c)}{\pi_e^{\theta'}(m \mid z'_c)}\right),
\end{equation}
where the partitioned probability $\pi^{\theta}_e(m \mid z)$ is again a product distribution over the parts in the partition, as in \eqref{eq:partitioned_effect}.

Taken together, the intrinsic information \eqref{eq:iie'} determines what cause or effect state the mechanism $m$ specifies. Its integrated information quantifies to what extent $m$ specifies its cause or effect in an irreducible manner. Again, $\varphi(m,Z)$ is a quantifier of irreducible existence.


\subsubsection*{Exclusion: Determining causal distinctions}

Finally, to comply with exclusion, a mechanism must select a definite effect purview, as well as a cause purview, out of a set of candidate purviews. Resorting again to the principle of maximal existence, the mechanism's effect purview and associated effect is the one having the maximum value of integrated information across all possible purviews $Z \subseteq S$ in state $z'_e(m,Z)$ \eqref{eq:effect_state}
\begin{equation}
\label{eq:Zstar}
z_e^*(m) = \argmax_{\{z'_e | Z \subseteq S\}} \varphi_e(m, Z = z'_e). 
\end{equation}
The integrated effect information of a mechanism $m$ within $S$ is then
\begin{equation}
\label{eq:phie}
 \varphi_e(m) := \varphi_e(m, z^*_e) = \max_{\{z'_e | Z \subseteq S\}} \varphi_e(m, Z = z'_e).   
\end{equation}

The integrated cause information $\varphi_c(m)$ and the maximally irreducible cause $z^*_c(m)$ are defined in the same way.
Based again on the principle of minimal existence, the irreducibility of the distinction specified by a mechanism is given by the minimum between its integrated cause and effect information
\begin{equation}
\label{eq:phid}
    \varphi_d(m) = \min \bigl(\varphi_c(m), \varphi_e(m)\bigr). 
\end{equation}

\subsubsection*{Determining the set of causal distinctions that are congruent with the system cause--effect state}

\begin{figure}[h!]
    \includegraphics[width=\columnwidth]{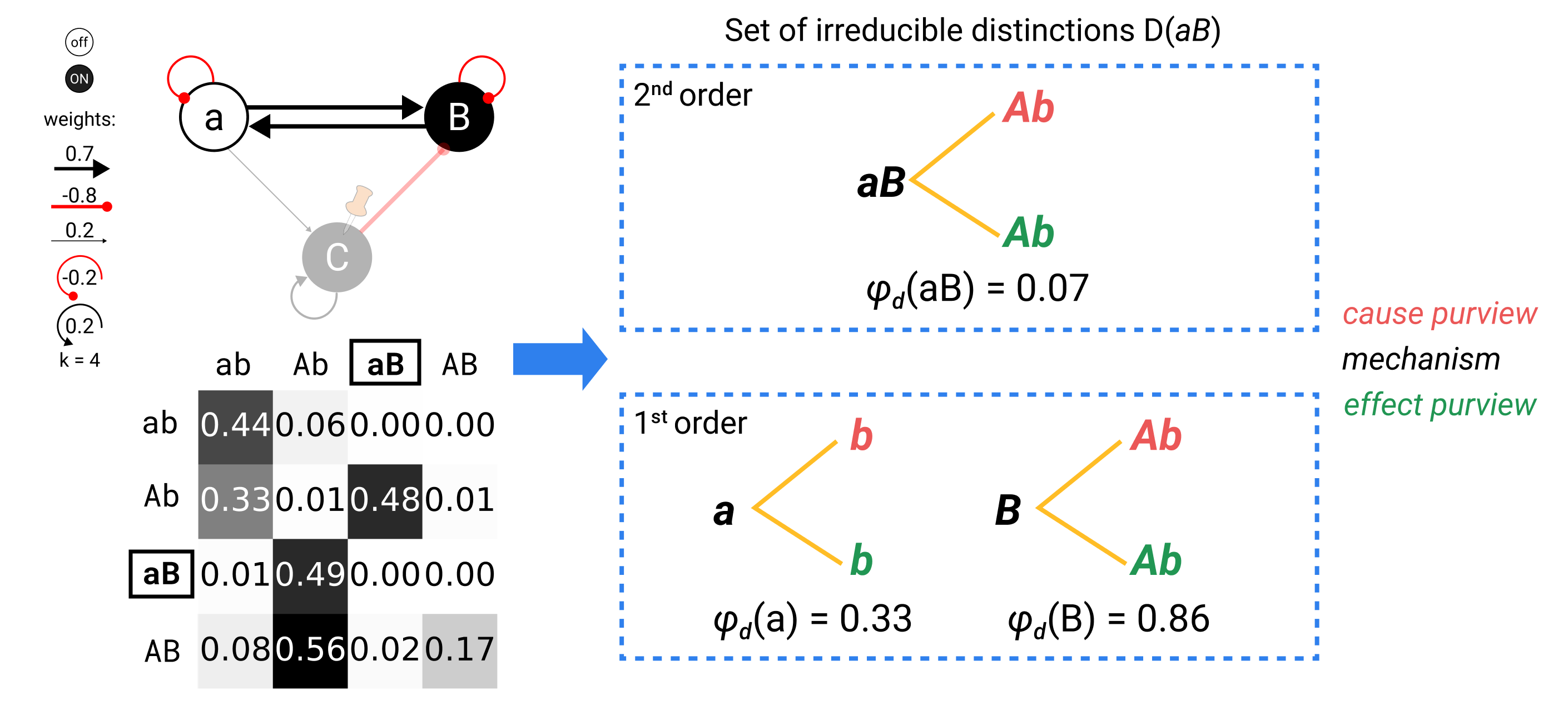}
    \caption{Composition and causal distinctions. Identifying the irreducible causal distinctions specified by a substrate in a state requires evaluating the specific causes and effects of every system subset. The candidate substrate is constituted of two interacting units $S = aB$ (see Fig. \ref{fig:SystemPhi}) and updates its state according to the depicted transition probability matrix. In addition to the two first-order mechanisms $a$ and $B$, the second-order mechanism $aB$ specifies its own irreducible cause and effect, as indicated by $\varphi_d > 0$.}
    \label{fig:HOdistinctions}
\end{figure}

As required by composition, unfolding the full cause--effect structure of the system $S$ in state $s$ requires assessing the irreducible cause--effect power of every subset of units within $S$ (Fig. \ref{fig:HOdistinctions}).
Any $m \subseteq s$ with $\varphi_d > 0$ specifies a candidate distinction $d(m) = (m, z^*, \varphi_d)$ \eqref{eq:distinction} within the system $S$ in state $s$. 
However, in order to contribute to the cause--effect structure of a system, distinctions must also comply with intrinsicality and information at the system level. This means that the cause--effect state they specify over subsets of the system ($z^* = \{z^*_{c}, z^*_{e}\}$) must be congruent with the cause--effect state specified over itself by the system as a whole $s'$.

We thus define the set of all causal distinctions within $S$ in state $s$ as
\begin{equation}
\label{eq:setD}
    \D(\T_S, s) = \{d(m) ~:~m \subseteq s, ~ \varphi_d(m) > 0, ~z^*_c(m) \subseteq s'_c, ~z^*_e(m) \subseteq s'_e \}.                          
\end{equation} 

Altogether, distinctions can be thought of as irreducible ``handles'' through which the system can take and make a difference to itself by linking an intrinsic cause to an intrinsic effect over subsets of itself. 
As components within the system, causal distinctions have no inherent structure themselves. Whatever structure there may be between the units that make up a distinction is not a property of the distinction but due to the structure of the system, and thus captured already by its compositional set of distinctions. 
Similarly, from an extrinsic perspective, one may uncover additional causes and effects, both within the system and across its borders, at either macro or micro grains. However, from the intrinsic perspective of the system causes and effects that are excluded from its cause--effect structure do not exist \cite{Albantakis2019cc, Grasso2021}.

\begin{figure}[h!]
    \includegraphics[width=11cm]{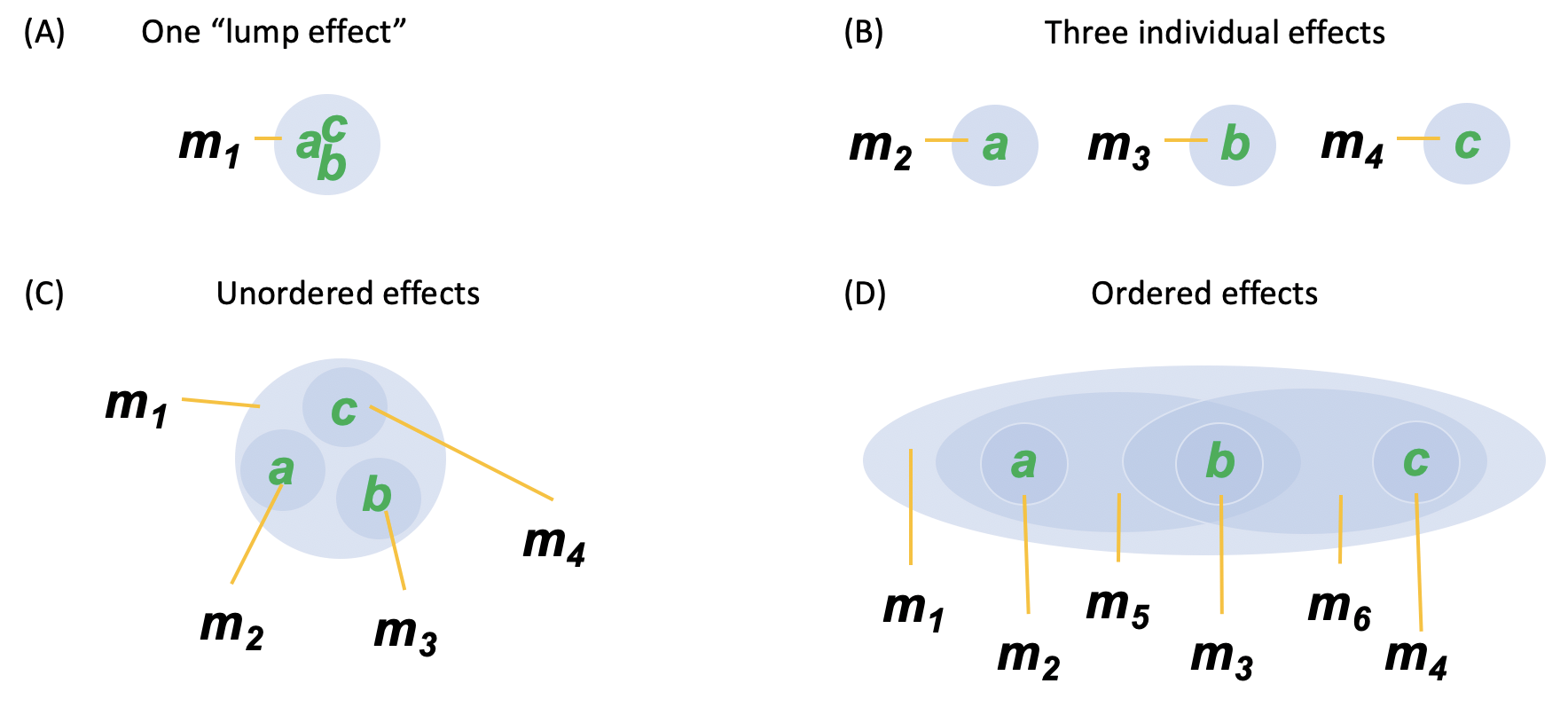}
    \caption{Composition of intrinsic effects. From the intrinsic perspective of the system, a specific cause or effect is only available to the system if it is selected by a causal distinction $d \in D(s)$. In (A), only the top-order effect is specified. From the intrinsic perspective, the system cannot distinguish the individual units. In (B), only first-order effects are specified. The system has no ``handle'' to select all three units together. (C) If both first- and third-order effects are specified, but no second-order effects, the system can distinguish individual units and select them together, but has no way of ordering them sequentially. (D) The system can distinguish individual units, select them altogether, as well as order them sequentially, in the sense that it has a handle for $ab$ and $bc$, but not $ac$. The ordering becomes apparent once the relations among the distinctions are considered (see below).}
    \label{fig:Space}
\end{figure}

For example, as shown in Fig. \ref{fig:Space}(A), a system may have a mechanism through which it specifies, in a maximally irreducible manner, the effect state of a triplet of units (\emph{e.g.}, $z^*_e = abc$, a third-order purview; again lowercase letters for units indicate state ``$-1$'', uppercase letters state ``$+1$''). However, if the system lacks a mechanism through which it can specify the effect state of single units, each taken individually (say, unit $a$, a first-order effect purview), then, from its intrinsic perspective, that unit does not exist as a single unit. By the same token, if the system can specify individually the state of unit $a$, $b$, and $c$, but lacks a way to specify irreducibly the state of $abc$ together, then, from its intrinsic perspective, the triplet $abc$ does not exist as a triplet (see Fig. \ref{fig:Space}(B)). Finally, even if the system can distinguish the single units $a$, $b$, and $c$, as well as the triplet $abc$, if it lacks handles to distinguish pairs of units such as $ab$ and $bc$, it cannot order units in a sequence.


\subsection*{Composition and causal relations}

Causal relations capture how the causes and/or effects of a set of distinctions within a complex overlap with each other. 
Just as a distinction specifies which units/states constitute a cause purview and the linked effect purview, a relation specifies which units/states correspond to which units/states among the purviews of a set of distinctions. 
Relations thus reflect how the cause--effect power of its distinctions is ``bound together'' within a complex. 
The irreducibility due to this binding of cause--effect power is measured by the relations' irreducibility ($\varphi_r > 0$).
Relations between distinctions were first described in \cite{Haun2019} (for differences with the initial presentation see \ref{A:Comparison}).

A set of distinctions $\bm{d} \subseteq D(s)$ is related if the cause, or effect, or both the cause and effect of each distinction $d \in \bm{d}$ overlap congruently over a set of shared units. Below we will denote the cause of a distinction $d$ as $z^*_c(d)$ and its effect as $z^*_e(d)$.
For a given set of distinctions $\bm{d} \subseteq D(s)$, there are potentially many ``relating'' sets of causes and/or effects $\bm{z}$ such that
\begin{equation}
\label{eq:relatingset}
    \bm{z} \; :\; \bm{z} \cap \{z^*_c(d), z^*_e(d)\} \neq \varnothing \;\; \forall d \in \bm{d},\; \bigcap_{z \in \bm{z}} z \neq \varnothing,\; |\bm{z}| > 1
\end{equation}
with maximal overlap

\begin{equation}
\label{eq:overlap}
    o^*(\bm{z}) = \bigcap_{z \in \bm{z}} z \neq \varnothing.
\end{equation}
All possible sets $\bm{z}$ specify unique aspects about a relation $r(\bm{d})$ and constitute the various ``faces'' of the relation.
The maximal overlap $o^*(\bm{z})$ \eqref{eq:overlap} is also called the ``face purview.'' 
Note that \eqref{eq:relatingset} includes the case $\bm{z} = \{z^*_c(d), z^*_e(d)\}$, which indicates a ``self-relation'' over the cause and effect of a single distinction $d \in \D(s)$.

A relation $r(\bm{d})$ thus consists of a set of distinctions $\bm{d} \subseteq D(s)$, with an associated set of faces $\bm{f}(\bm{d}) = \{ f(\bm{z})\}_{\bm{d}}$ and irreducibility $\varphi_r > 0$,

\begin{equation}
\label{eq:relation}
    r(\bm{d}) = \Big(\bm{d}, \bm{f}(\bm{d}), \varphi_r\Big).
\end{equation}
A relation that binds together $h = |\bm{d}|$ distinctions is a $h$-degree relation. 
A relation face $f(\bm{z}) \in \bm{f}(\bm{d})$ consists of a set of causes and effects $\bm{z}$ (as in \refeq{eq:relatingset}), with associated face purview $o^*(\bm{z}) \eqref{eq:overlap}$

\begin{equation}
\label{eq:faces}
    f(\bm{z}) = \Big(\bm{z}, o^*(\bm{z})\Big).
\end{equation}
A relation face over $k = |\bm{z}|$ purviews is a $k$-degree face.
Because $\bm{z}$ may include either the cause, or the effect, or both the cause and effect of a distinction $d \in \bm{d}$, a relation $r(\bm{d})$ with $|\bm{d}| > 1$ may comprise up to $3^{|\bm{d}|}$ faces.
If a set of distinctions $\bm{d} \in D(s)$ does not overlap congruently, it is not related (in that case $o^*(\bm{z}) = \varnothing$ for all possible $f(\bm{z}) \in \bm{f}(\bm{d})$). 

Causal relations inherit existence from the cause--effect power of the distinctions that compose them. They inherit intrinsicality because the causes and effects that compose their faces are specified within the substrate. 
Moreover, relations are specific because the joint purviews of their faces must be congruent for all causes and effects $z^* \in \bm{z}$.
Note that relation purviews are necessarily congruent with the overall cause and effect state specified by the system as a whole, because the causes and effects of the distinctions composing a relation must themselves be congruent.

The irreducibility of a causal relation is measured by ``unbinding'' distinctions from their joint purviews, taking into account all faces of the relation. Distinctions $d \in D(s)$ are already established as maximally irreducible components, characterized by their value of integrated information $\varphi_d$. 
To assess the irreducibility of a relation, we thus assume that the integrated information $\varphi_d$ of a distinction is distributed uniformly across unique cause and effect purview units, such that 
\begin{equation}
\label{eq:rallocation}
    \frac{\varphi_d}{|z^*_c(d)\cup z^*_e(d)|}
\end{equation}
is the average irreducible information $\varphi_d$ per unique purview unit for an individual distinction $d \in \bm{d}$ with cause--effect state $z^*(d) =  \{z^*_c(d), z^*_e(d)\}$ (congruent units on the cause and effect side count as one, while incongruent units are counted separately). 

Since distinctions are related by constraining common units (though they may constrain them in different ways), the effect of ``unbinding'' a distinction must be proportional to the number of units jointly constrained in the relation, \emph{i.e.} the number of unique units over the joint purviews of all faces in the relation: 
\begin{equation}
\label{eq:roverlapunion}
    \left|\bigcup_{f\in\bm{f}(\bm{d})} \; o^*_f \;\right|.
\end{equation}
This union of the face purviews $o^*_f$ is also called the ``relation purview'' or the ``joint purview'' of the relation. While any partition of one or more distinctions from the relation will ``unbind'' the set of distinctions $\bm{d}$, by the principle of minimal existence, a relation can only be as irreducible as the minimal amount of integrated information specified by any one distinction in the relation. 
Therefore, the relation integrated information $\varphi_r(\bm{d})$ is defined as

\begin{equation}
\label{eq:phir}
    \varphi_r(\bm{d}) = \left|\bigcup_{f\in\bm{f}(\bm{d})} \; o^*_f \;\right| \min_{d \in \bm{d}} \; \frac{\varphi_d}{|z^*_c(d)\cup z^*_e(d)|}.
\end{equation}
Defining $\varphi_r$ in this way guarantees that the integrated information of a relation cannot exceed the integrated information of its weakest distinction. A maximal relation is thus a relation in which the cause and effect of each distinction is fully overlapped by other distinctions in the relation (in that case, $\varphi_r = \min_{d\in\bm{d}} \varphi_d$). 
Note also that a relation satisfies exclusion in that its integrated information is naturally maximized over the maximally congruent overlap $o^*_f$ for each relation face \eqref{eq:overlap} (taking subsets of these overlaps could only reduce the integrated information of the relation). 

\begin{figure}[h!]
    \includegraphics[width=10cm]{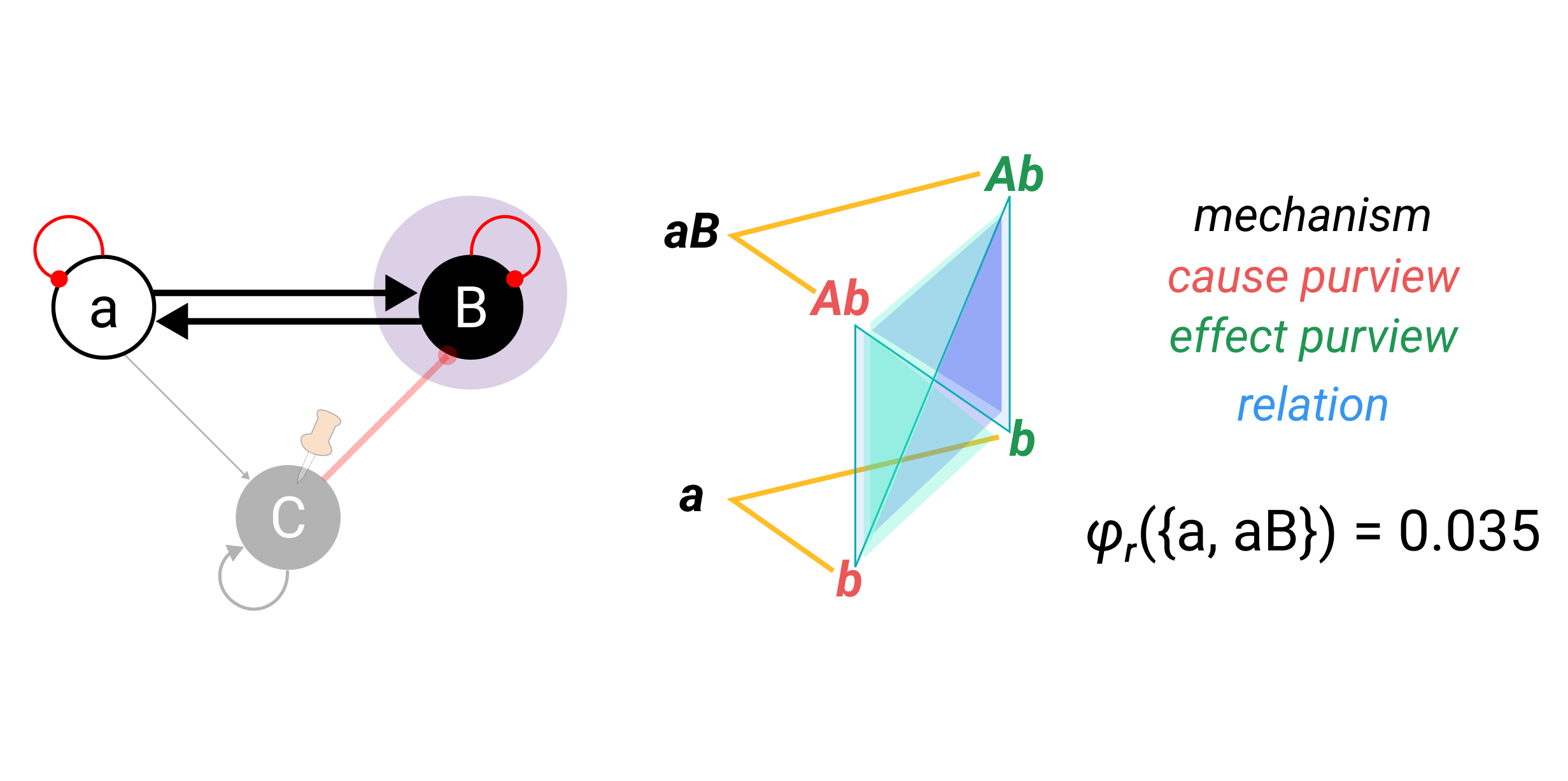}
    \caption{Composition and causal relations. Relations between distinctions specify joint causes and/or effects. The two distinctions $d(a)$ and $d(aB)$ each specify their own cause and effect. In this example, their cause and effect purviews overlap over the unit $b$ and are congruent, which means that they all specify $b$ to be in state ``-1''. The relation $r(\{a, ab\})$ thus binds the two distinctions together over the same unit. Relation faces are indicated by the blue lines and surfaces between the distinctions' causes and/or effects. Because all four purviews overlap over the same unit, all nine possible faces exist. Note that the fact that the two distinctions overlap irreducibly can only be captured by a relation and not by a high-order distinction.}
    \label{fig:RelationsAB}
\end{figure}

In summary, just as distinctions link a cause with an effect, relations bind various combinations of causes and effects that are congruent over the same units, \emph{i.e.}, constraining those units to be in the same state (Fig. \ref{fig:RelationsAB}). 
And just as a distinction captures the irreducibility of an individual cause--effect linked by a mechanism, a relation captures the irreducibility of a set of distinctions bound by the joint purviews of their causes and/or effects.

For a set of distinctions $\D$, we define the set of all relations among them as 
\begin{equation}
    {R}(D) = \{r(\bm{d}) : \varphi_r(\bm{d}) > 0 \}, \; \forall \bm{d} \subseteq D.
\end{equation}
In practice, the total number of relations and their $\sum_{\bm{R(D)}} \varphi_r$ can be determined analytically for a given set of distinctions $D$, which greatly reduces the necessary computations (see \ref{A:AnalyticalRelations}). 
Together, a set of distinctions $D$ and its associated set of relations ${R}(D)$ compose a cause--effect structure.
\begin{figure}[h!]
    \includegraphics[width=10cm]{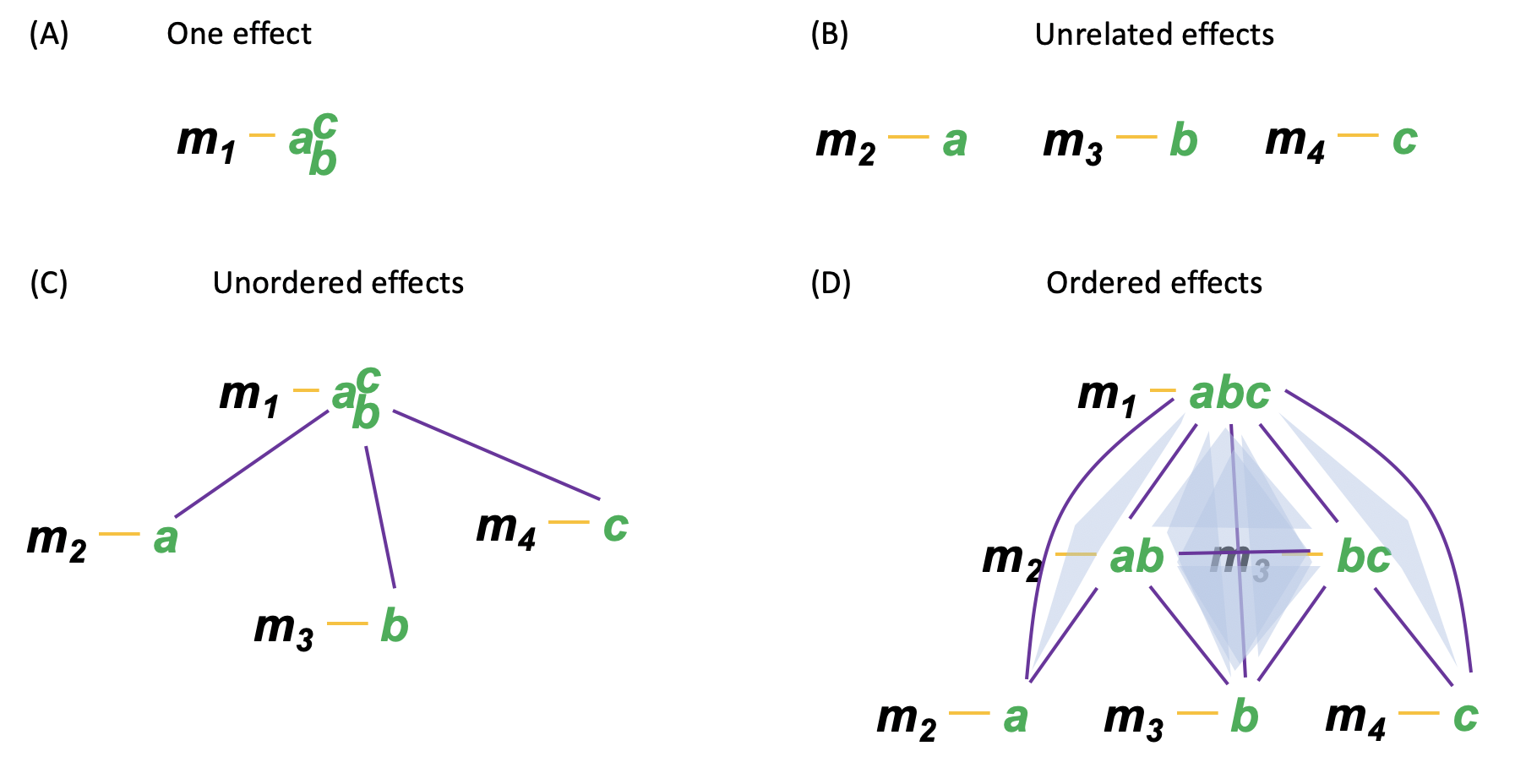}
    \caption{Structuring of intrinsic effects by relations. (A) A single undifferentiated effect has no relations. (B) Likewise, there are no relations among multiple non-overlapping effects. (C) The set of three first-order effects and one third-order effect supports three relations, which bind the effects together. (D) The set of first, second, and third-order effects supports a large number of relations (10 2-relations (between two effects), 6 3-relations, and 1 4-relation), which bind the effects in a structure that is ordered sequentially.}
    \label{fig:SpaceRelations}
\end{figure}

\subsection*{Cause--effect structures and $\varPhi$-structures}
A cause--effect structure is defined as the union of the distinctions specified by a substrate and the relations binding them together:
\begin{equation}
    C(D) = D \cup {R}(D).
\end{equation}
The cause--effect structure specified by a maximal substrate---a complex---is also called a $\varPhi$-structure: 

\begin{adjustwidth}{-1.5in}{0in}
\begin{equation}
    C(\T_{S^*}, s^*) = \Bigg\{ \Big\{d(m) = \{m, z^*, \varphi_d\} \in \D(\T_{S^*}, s^*)\Big\}
    ~\bigcup~\Big\{r(\bm{d}) = \{\bm{d}, \bm{f}(d), \varphi_r\} \in R\bigl(D(\T_{S^*}, s^*)\bigr)\Big\}\Bigg\}.
\end{equation}
\end{adjustwidth}

The sum of the values of integrated information of a substrate's distinctions and relations, called $\varPhi$ (``big phi'') corresponds to the \textit{structured information} of the $\varPhi$-structure,

\begin{equation}
    \varPhi(\T_{S^*}, s^*) = \sum_{C(\T_{S^*}, s^*)} \varphi. 
\end{equation}

In conclusion, a maximal substrate or complex is a set of units $S^* = s^*$ that satisfies all of IIT's postulates: it has to have cause--effect power that is intrinsic, specific, irreducible, definite, and structured. 
By IIT, a complex $S^*$ does not exist as such, but exists ``unfolded'' into its $\varPhi$-structure, with all the causal distinctions and relations that compose it. In other words, a substrate is what can be observed and manipulated ``operationally'' from the extrinsic perspective. From the intrinsic perspective, what truly exists is a complex with all its causal powers unfolded---an \textit{intrinsic entity} that exists for itself, absolutely, rather than relative to an external observer.

According to the explanatory identity of IIT, an experience is identical to the $\varPhi$-structure of an intrinsic entity: every property of the experience should be accounted for by a corresponding property of the $\varPhi$-structure, with no additional ingredients.
If a system $S$ in state $s$ is a complex, then its $\varPhi$-structure corresponds to the quality of the experience of $S$ in state $s$, while its $\varPhi$ value corresponds to its quantity---in other words, the nature and amount of intrinsic content. 

\pagebreak
\section*{Results and discussion}

In this section, we apply the mathematical framework of IIT 4.0 to several example systems. 
The goal is to illustrate three critical implications of IIT's postulates: 
\begin{enumerate}
    \item \textbf{Consciousness and connectivity: }how the way units interact determines whether a substrate can support a $\varPhi$-structure of high $\varPhi$ 
    \item \textbf{Consciousness and activity:} how changes in the activity of a substrate's units change $\varPhi$-structures 
    \item \textbf{Consciousness and functional equivalence:} how substrates that are functionally equivalent may not be equivalent in terms of their $\varPhi$-structures, and thus in terms of consciousness 
\end{enumerate}

The following examples will feature very simple networks constituted of binary units $U_i \in U$ with $\Omega_{U_i} = \{-1,1\}$ for all $U_i$ and a logistic (sigmoidal) activation function
\begin{equation}
\label{eq:sig}
    p(U_{i,t} = 1 \mid u_{t - 1}) = \frac{1}{1 + \exp\left(-k\sum_{j = 1}^n w_{j,i}u_{j, t-1}\right)},
\end{equation}
where 
\begin{equation}
    \sum_{j = 1}^n w_{j,i} = 1 ~ \forall ~ i.
\end{equation}

In Eq. \eqref{eq:sig}, the parameter $k$ defines the slope of the logistic function and allows one to adjust the amount of noise or determinism in the activation function (higher values signify a steeper slope and thus more determinism). The units $U_i$ can thus be viewed as noisy linear threshold units with weighted connections among them.

As in Figs. \ref{fig:SystemPhi} and \ref{fig:HOdistinctions}, units denoted by uppercase letters are in state `$1$' (ON, depicted in black), units denoted by lowercase letters are in state `$-1$' (OFF, depicted in white).
Cause--effect structures are illustrated as geometrical shapes projected into 3D space (Fig. \ref{fig:architecture}). 
Distinctions are depicted as mechanisms (black labels) tying a cause (red labels) and an effect (green labels) through a link (orange edges, thickness indicating $\varphi_{d}$). 
Relation faces of second- and third-degree relations are depicted as blue edges or triangular surfaces between the causes and effects of the related distinctions. While edges always bind pairs of distinctions (a second-degree relation), triangular surfaces may bind the causes and effects of two or three distinctions (second- or third-degree relation). Relations of higher degrees are not depicted.

All examples were computed using the ``iit-4.0'' feature branch of PyPhi \cite{Mayner2018}. This branch is currently under development and will be available in the next official release of the software. An example notebook is \href{https://colab.research.google.com/github/wmayner/pyphi/blob/feature/iit-4.0/docs/examples/IIT_4.0_demo.ipynb}{available here}. 

\subsection*{Consciousness and connectivity}
The first set of examples highlights how the organization of connections among units impacts the ability of a substrate to support a cause--effect structure with high structured information (high $\varPhi$). 
Fig. \ref{fig:architecture} shows five systems, all in the same state $s = \textit{Abcdef}$ with the same number of units, but with different connectivity among the units.

\begin{figure}[p]
    \includegraphics[width=12cm]{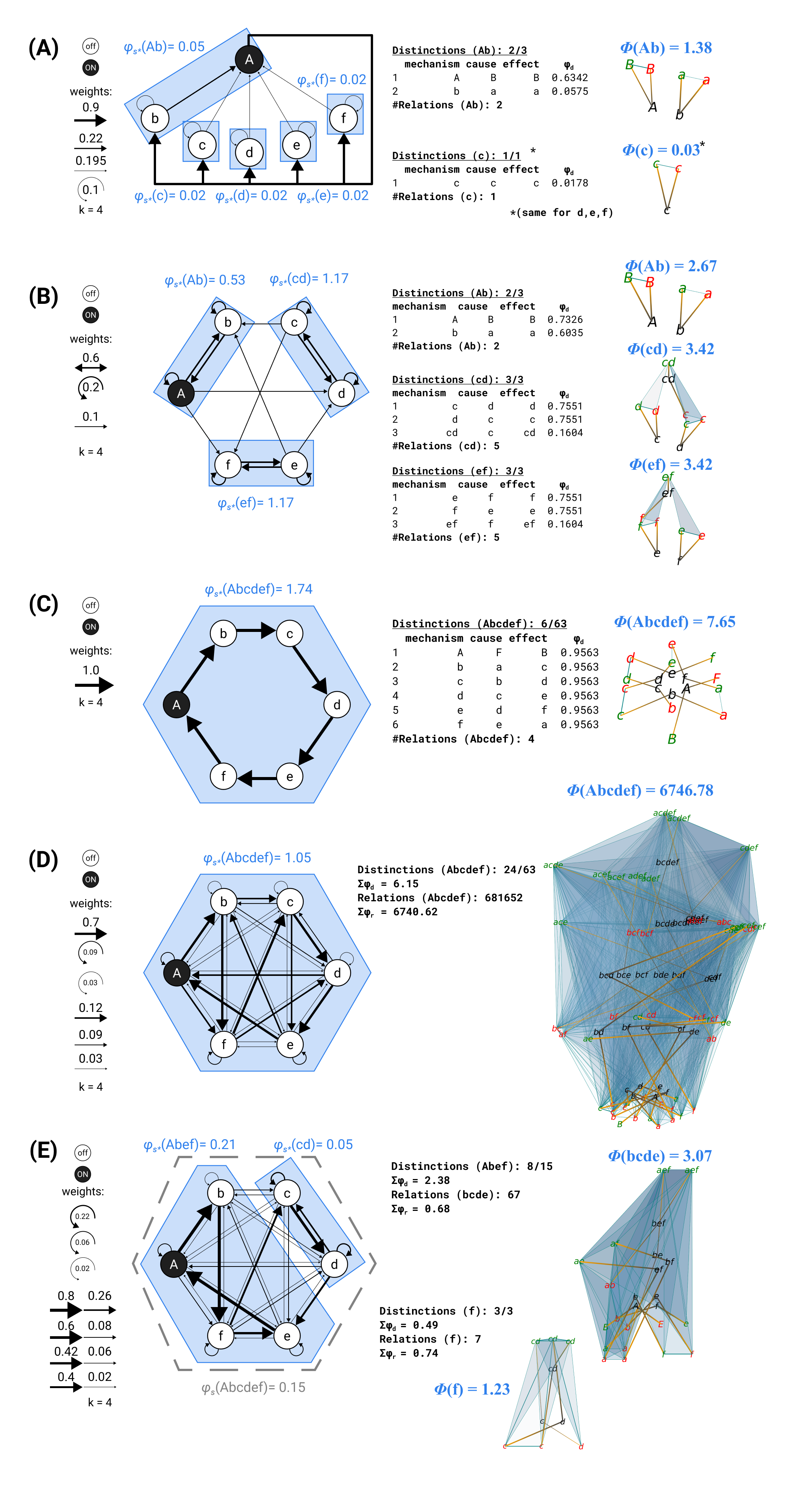} 
\end{figure}

\clearpage 
\begingroup
\captionof{figure}{\textbf{Causal analysis of various network architectures.} Each panel shows the network's causal model and weights on the left. Blue regions indicate complexes with their respective $\varphi_s$ values. In all networks, $k = 4$ and the state is $Abcdef$. The $\varPhi$-structure(s) specified by the network's complexes are illustrated below or to the right (with only second- and third-degree relation faces depicted) with a list of their distinctions for smaller systems and their $\sum \varphi$ values for those systems with many distinctions and relations. All integrated information values are in ibits. (A) A degenerate network in which unit $A$ forms a bottleneck with redundant inputs from and outputs to the remaining units. The first-maximal complex is $Ab$, which excludes all other subsets with $\varphi_s > 0$ except for the individual units $c$, $d$, $e$, and $f$. (B) The modular network condenses into three complexes along its fault lines (which exclude all subsets and supersets), each with a maximal $\varphi_s$ value, but low $\varPhi$, as the modules each specify only two or three distinctions and at most five relations. (C) A directed cycle of six units forms a six-unit complex with $\varphi_s = 1.74$ ibits, as no other subset is integrated. However, the $\varPhi$-structure of the directed cycle is composed of only first-order distinctions and few relations. (D) A specialized lattice also forms a complex (which excludes all subsets), but specifies 24 first- and high-order distinctions, with many relations ($>6*10^5$) among them. Its $\varPhi$ value is $6747$ ibits, much larger than the structured information of the complexes that exist in the other networks. (E) A slightly modified version of the specialized lattice in which the first-maximal complex is $Abef$. The full system is not maximally irreducible and is excluded as a complex, despite its positive $\varphi_s$ value (indicated in gray).} 
\label{fig:architecture}
\endgroup

\subsubsection*{Degenerate systems, indeterminism, and specificity} 
Fig. \ref{fig:architecture}A shows a network with medium indeterminism ($k = 4$) and high degeneracy, due to the fact that unit $A$ forms a ``bottleneck'' with inputs and outputs to and from the remaining units. 
The network condenses into one complex of two units $Ab$ and four complexes corresponding to the individual units $c$, $d$, $e$, and $f$ (also called ``monads'').  

The causes and effects of the causal distinctions for the two types of complexes are shown in the middle, and the corresponding cause--effect structures are illustrated on the right. 
In this case, degeneracy (coupled with indeterminism) undermines the ability of the maximal substrate to grow in size, which in turn limits the richness of the $\varPhi$-structure that can be supported. 
Because of the bottleneck architecture, the current state of candidate system $Abcdef$ has many possible causes and effects, leading to an exponential decrease in selectivity (the conditional probabilities of cause and effect states).
This dilutes the value of intrinsic information ($\ii$) for larger subsets of units, which in turn reduces their value of system integrated information $\varphi_s$. 
Consequently, the maximal substrates are small, and their $\varPhi$ values are necessarily low. 

This example suggests that to grow and achieve high values of $\varPhi$, substrates must be constituted of units that are specialized (low degeneracy) and interact very effectively (low indeterminism). 
Notably, the organization of the cerebral cortex, widely considered as the likely substrate of human consciousness, is characterized by extraordinary specialization of neural units at all levels \cite{Kanwisher2010, Ponce2019, Khosla2022}. Moreover, if the background conditions are well controlled, neurons are thought to interact in a highly reproducible, nearly deterministic manner \cite{Mainen1995, Hires2015, Nolte2019}.    

\subsubsection*{Modular systems, fault lines, and irreducibility}

Fig. \ref{fig:architecture}B shows a network comprising three weakly interconnected modules, each having two strongly connected units ($k = 4$). 
In this case, the weak inter-module connections are clear fault lines.
Properly normalized, partitions along these fault lines separating modules yield values of $\varphi_s$ that are much smaller than those yielded by partitions that cut across modules. 
As a consequence, the 6-unit system condenses into three complexes (\textit{Ab}, \textit{cd}, and \textit{ef}), as determined by their maximal $\varphi_s$ values. 
Again, because the modules are small, their $\varPhi$ values are low. 
Intriguingly, a brain region such as the cerebellum, whose anatomical organization is highly modular, does not contribute to consciousness \cite{Lemon2010, Yu2015}, even though it contains several times more neurons than the cerebral cortex (and is indirectly connected to it). 

Note that fault lines can be due not just to neuroanatomy but also to neurophysiological factors. 
For example, during early slow-wave sleep, the dense interconnections among neuronal groups in cerebral cortical areas may break down, becoming causally ineffective due to the bistability of neuronal excitability. This bistability, brought about by neuromodulatory changes \cite{Steriade1993}, is associated with the loss of consciousness \cite{Pigorini2015}.

\subsubsection*{Directed cycles, structural sparseness, and composition}

Fig. \ref{fig:architecture}C shows a directed cycle in which six units are unidirectionally connected with weight $w = 1.0$ and $k=4$.
Each unit copies the state of the unit before it, and its state is copied by the unit after it, with some indeterminism. 
The copy cycle constitutes a 6-unit complex with a maximal $\varphi_s = 1.74$ ibits. 
However, despite the ``large'' substrate, the $\varPhi$-structure it specifies has low structured information ($\varPhi = 7.65$). 
This is because the system's $\varPhi$-structure is composed exclusively of first-order distinctions, and consequently of a small number of relations. 

Highly deterministic directed cycles can easily be extended to constitute large complexes, being more irreducible than any of their subsets. 
However, the lack of cross-connections (``chords'' in graph-theoretic terms) greatly limits the number of components of the $\varPhi$-structures the complexes specify, and thus their structured information ($\varPhi$). (Note also that increasing the number of units that constitute the directed cycle would not change the amount of $\varphi_s$ specified by the network as a whole.)

The brain is rich in partially segregated, directed cycles, such as those originating in cortical areas, sequentially reaching stations in the basal ganglia and thalamus, and cycling back to cortex \cite{Middleton2000, Foster2021}. These cycles are critical for carrying out many cognitive and other functions, but they do not appear to contribute directly to experience \cite{Tononi2016}.

\subsubsection*{Specialized lattices and $\varPhi$-structures with high structured information}

Fig. \ref{fig:architecture}D shows a network consisting of six heterogeneously connected units---a ``specialized'' lattice, again with $k=4$. 
While many subsystems within the specialized network have positive values of system integrated information $\varphi_s$ (not shown), the full 6-unit system is the maximal substrate (excluding all its subsets from being maximal substrates). 
Out of 63 possible distinctions, the $\varPhi$-structure comprises 24 distinctions with causes and effects congruent with the system's maximal cause--effect state. Consequently, the full 6-unit system also specifies a much larger number of causal relations compared to the copy loop system.

Preliminary work (not shown, \cite{Fujii}) indicates that lattices of specialized units, implementing different input--output functions, but partially overlapping in their inputs (receptive field) and outputs (projective fields), are particularly well suited to constituting large substrates that unfold into extraordinarily rich $\varPhi$-structures. 
The number of distinctions specified by an optimally connected, specialized system is bounded above by $2^n-1$, and that of the relations among as many distinctions is bounded by $2^{(2^n-1)}-1$.The structured information composing such structures is correspondingly large \cite{Zaeemzadeh}.

In the brain, many portions of posterior cortex appear to constitute similarly organized lattices of specialized units, which makes the posterior cortex and similarly organized areas a plausible candidate for the substrate of human consciousness \cite{Tononi2016, Boly2017, Haun2019, Watakabe2021}. Note that directed cycles originating and ending in such lattices typically remain excluded from the first-maximal complex because minimal partitions across such cycles yield a much lower value of $\varphi_s$ compared to minimal partitions across large lattices \cite{Fujii}.     

\subsubsection*{Near-maximal substrates, extrinsic entities, and exclusion} 
Finally, Fig. \ref{fig:architecture}E shows a network of six units, four of which ($Abef$) constitute a specialized lattice that corresponds to the first complex. Though integrated, the full set of 6 units happens to be slightly less irreducible than one of its 4-unit subsets ($\varphi_s = 0.15$). 
From the extrinsic perspective, the 6-unit system undoubtedly behaves as a highly integrated whole (nearly as much as its 4-unit subset), one that could produce complex input--output functions due to its rich internal structure. 
From the intrinsic perspective of the system, however, only the 4-unit subset satisfies all the postulates of existence, including maximal irreducibility (accounting for the definite nature of experience).
In this example, the remaining units form a second complex with low $\varphi_s$ and serve as background conditions for the first complex. 

A similar situation may occur in the brain.
The brain as a whole is undoubtedly integrated (not to mention that it is integrated with the body as a whole), and neural ``traffic'' is heavy throughout. 
However, its anatomical organization may be such that a subset of brain regions, arranged in pyramids of grids and primarily located in posterior cortex, may achieve a much higher value of integrated information than any other subset. Those regions would then constitute the first complex (the ``main complex,'' \cite{Tononi2016}), and the remaining regions might condense into a large number of much smaller complexes. 
\newline

Taken together, the examples in Fig. \ref{fig:architecture} demonstrate that the connectivity among the units of a system has a strong impact on what set of units can constitute a complex and, thus, on the structured information it can specify. 
The examples also demonstrate the role played by the various requirements that must be satisfied by a substrate of consciousness: existence (causal power), intrinsicality, specificity, irreducibility, maximal irreducibility (exclusion), and composition (structure). 

\subsection*{Consciousness and activity: active, inactive, and inactivated units}
A substrate exerts cause--effect power by being in its current state. 
For the same substrate, changing the state of even one unit may have major consequences on the distinctions and relations that compose its $\varPhi$-structure: many may be lost, or gained, and many may change their value of irreducibility ($\varphi_d$ and $\varphi_r$). 

\begin{figure}[h!]
    \includegraphics[width=10.9cm]{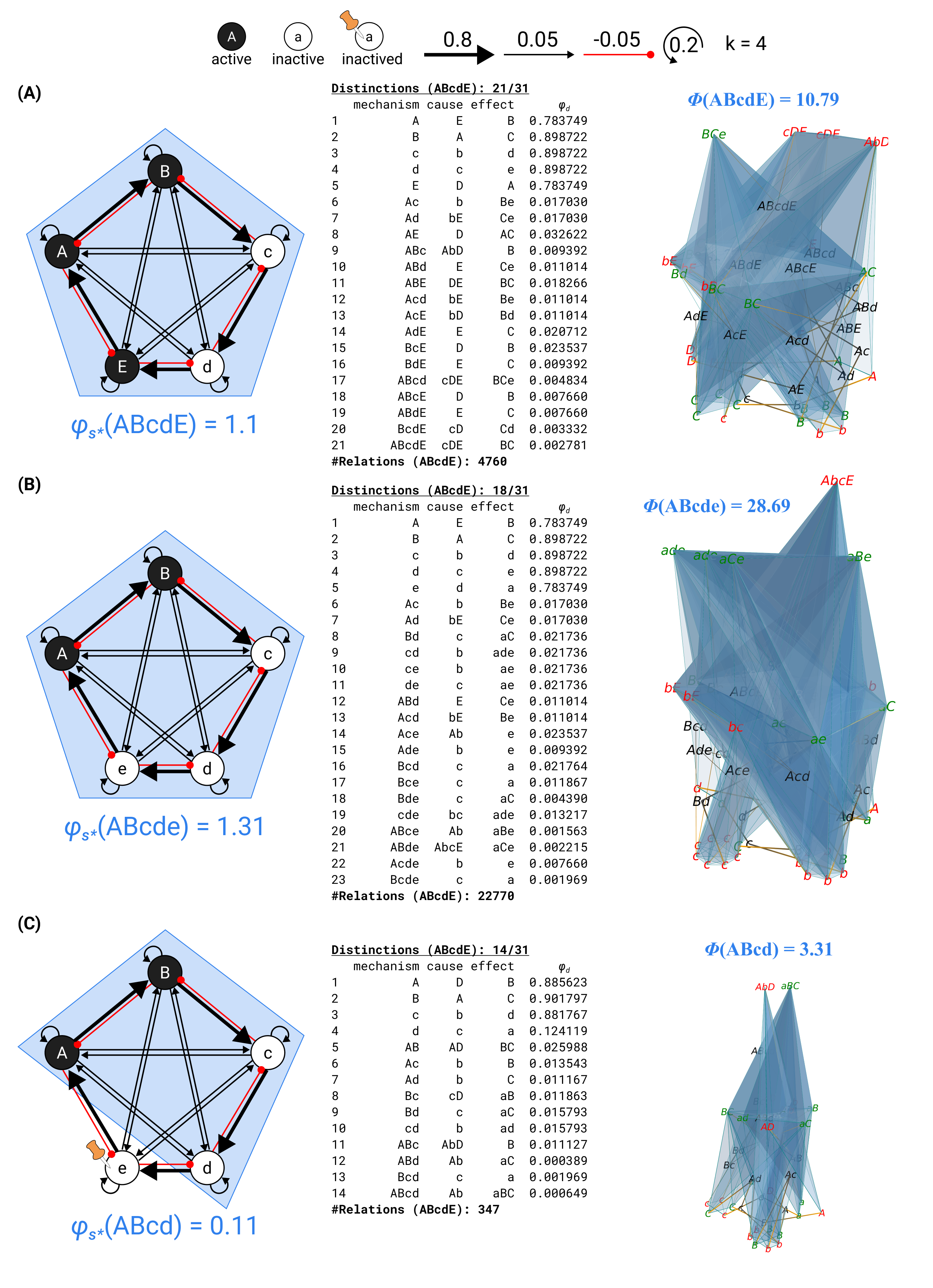} 
    \caption{\textbf{Causal analysis of the same system with one of its units set to active, inactive, or inactivated.} In all panels, the same causal model and weights are shown on the left, but in different states. For all networks $k = 4$. The set of distinctions $\D(s)$, their causes and effects, and their $\varphi_d$ values are shown in the middle. The $\varPhi$-structure specified by the network's complex is illustrated on the right (again with only second- and third-degree relation faces depicted). All integrated information values are in ibits. (A) The system in state $ABcdE$ is a complex with 21 out of 31 distinctions and $\varPhi = 10.79$. (B) The same system in state $ABcde$, where unit $E$ is inactive (``OFF'') also forms a complex with a similar number of distinctions, but a somewhat higher $\varPhi$ value due to a higher number of relations between distinctions. In addition, the system's $\varPhi$-structure differs from that in (A), as the system now specifies a different set of compositional causes and effects. (C) If instead of being inactive, unit $E$ is inactivated (permanently fixed into the ``OFF'' state), the inactivated unit cannot contribute to the complex or $\varPhi$-structure anymore. The complex is now constituted of four units ($ABcd$), with only 14 distinctions and significantly reduced structured information ($\varPhi = 3.31$).}
    \label{fig:state}
\end{figure}

Fig. \ref{fig:state} shows a network of five binary units that interact through excitatory and inhibitory connections (weights indicated in the figure). The system is initially in state $s = ABcdE$ (Fig. \ref{fig:state}A) and is a maximal substrate with $\varphi_s = 1.1$ ibits and a $\varPhi$-structure composed of 21 distinctions and their 4760 relations.  

If we change the state of unit $E$ from ON to OFF (in neural terms, the unit becomes inactive), the distinctions that the unit contributes to when ON, as well as the associated relations, may change (Fig. \ref{fig:state}B). 
In the case illustrated by the Figure, what changes are the purviews and irreducibility of several distinctions and associated relations, the number of distinctions and $\varphi_s$ change only slightly, while the number of relations is considerably higher, leading to a larger $\varPhi$ value.
In other words, what a single unit contributes to intrinsic existence is not some small ``bit'' of information. Instead, a unit contributes an entire sub-structure, composed of a very large number of distinctions and relations. 
The set of distinctions to which a subset of units contributes as a mechanism, either alone or in combination with other units, together with their associated relations, is called a mechanism \textit{$\varPhi$-fold}. 
With respect to the neural substrate of consciousness in the brain, this means that even a change in the state of a single unit is typically associated with a change in an entire $\varPhi$-fold within the overall $\varPhi$-structure, with a corresponding change in the structure of the experience.

In Fig. \ref{fig:state}, we see what happens if unit $E$, instead of just turning inactive (OFF) is \textit{inactivated} (abolishing its cause--effect power because it no longer has any counterfactual states). In this case, all the distinctions and relations to which that unit contributes as a mechanism cease to exist (its mechanism $\varPhi$-fold collapses). Moreover, all the distinctions and relations to whose purviews that unit contributes---its purview $\varPhi$-fold---also collapse or change. 
The complex also shrinks because it cannot include that unit. 
With respect to the neural substrate of consciousness, this means that while an inactive unit contributes to a different experience, an inactivated unit ceases to contribute to experience altogether. 
The fundamental difference between inactive and inactivated units leads to the following corollary of IIT: unlike a fully inactivated substrate which, as would be suspected, cannot support any experience, an inactive substrate can. If a maximal substrate is in working order and specifies a large $\varPhi$-structure, it will support a highly structured experience, such as the experience of empty space \cite{Haun2019} or the feeling of ``pure presence'' \cite{Boly_PurePresence}.   

\subsection*{Consciousness and functional equivalence: being is not doing}

By the intrinsicality postulate, the $\varPhi$-structure of a complex depends on the causal interactions between system subsets, not on the system's interaction with its environment. In general, different physical systems with different internal causal structure may perform the same input--output functions. 

\begin{figure}[h!]
    \hspace*{-3cm} 
    \captionsetup{margin={-3cm, 0cm}}
    \includegraphics[width=\columnwidth+3cm]{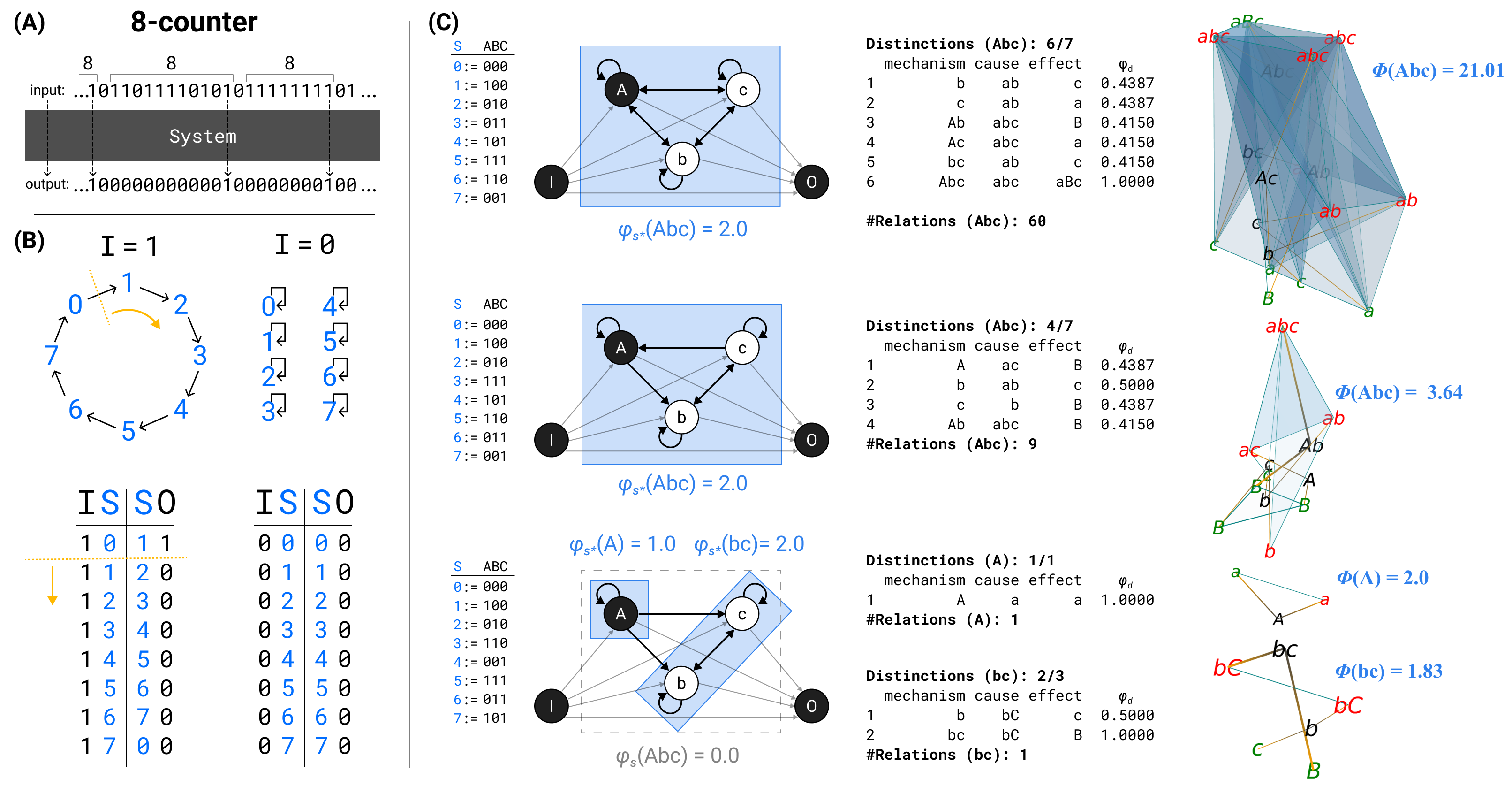}
    \caption{\textbf{Functionally equivalent networks with different $\varPhi$-structures.} (A) The input--output function realized by three different systems (shown in (C)): a count of eight instances of input $I=1$ leads to output $O=1$. (B) The global state-transition diagram is also the same for the three systems: if $I=0$, the systems will remain in their current global state, labeled as 0-7; if $I=1$, the systems will move one state forward, cycling through their global states, and activate the output if $S=0$. (C) Three systems constituted of three binary units but differing in how the units are connected and interact. As a consequence, the one-to-one mapping between the 3-bit binary states and the global state labels differ. However, all three systems initially transition from $000$ to $100$ to $010$. 
    Analyzed in state $100$, the first system (top) turns out to be a single complex that specifies a $\varPhi$-structure with six distinctions and many relations, yielding a high value of $\varPhi$. The second system (middle) is also a complex, with the same $\varphi_s$ value, but it specifies a $\varPhi$-structure with fewer distinctions and relations, yielding a lower value of $\varPhi$. Finally, the third system (bottom) is reducible ($\varphi_s = 0$) and splits into two smaller complexes (entities) with minimal $\varPhi$-structures and low $\varPhi$.}
    \label{fig:funcEquivalence}
\end{figure}

Fig. \ref{fig:funcEquivalence} shows three simple deterministic systems with binary units (here the ``OFF'' state is $0$, and ``ON'' is $1$) that perform the same input--output function, treating the internal dynamics of the system as a black box. 
The function could be thought of, for example, as an electronic tollbooth ``counting 8 valid coins'' (8 times input $I=1$) before opening the gate \cite{Hanson2021}.
Each system receives one binary input ($I$) and has one binary output ($O$). The output unit switches ``ON'' on a count of eight positive inputs $I = 1$ (when the global state with label `0' is reached in the cycle), upon which the system resets (Fig. \ref{fig:funcEquivalence}A).

In addition to being functionally equivalent in their outward behavior, the three systems share the same internal global dynamics, as their internal states update according to the same global state-transition diagram (Fig. \ref{fig:funcEquivalence}B). Given an input $I = 1$, the system updates its state, cycling through all its 8 global states (labeled 0--7) over 8 updates. For an input of $I=0$, the system remains in its present state. 
Moreover, all three systems are constituted of three binary units whose joint states map one-to-one onto the systems' global state labels (0--7). However, the mapping is different for different systems (Fig. \ref{fig:funcEquivalence}C, left). This is because the internal binary update sequence depends on the interactions among the internal units \cite{Albantakis2019cc, Hanson2021}, which differ in the three cases, as can easily be determined through manipulations and observations.  

For consistency in the causal analysis, in all three cases, the global state ``0'' that activates the output unit if $I = 1$ is selected such that it corresponds to the binary state ``all OFF'' ($000$), which is followed by $1 := 100$ and $2 := 010$. Also, the $\varPhi$-structure of each system is unfolded in state $1 := 100$ in all three cases.

Despite their functional equivalence and equivalent global dynamics, the systems differ in how they condense into complexes and in the cause--effect structures they specify. 

As shown in Fig. \ref{fig:funcEquivalence}C, the first system forms a 3-unit complex with a relatively rich $\varPhi$-structure ($\varPhi = 21.01$ ibits). While the second system also forms a 3-unit complex with the same $\varphi_s = 2$ ibits, it specifies a completely different set of distinctions and has much lower structured information ($\varPhi = 3.64$ ibits). 

Finally, the third system is reducible ($\varphi_s = 0$ ibits)---in this case, because there are only feed-forward connections from unit $A$ to units $B$ and $C$---and it condenses into two complexes with small $\varPhi$-structures.

These examples illustrate a simple scenario of functional equivalence of three systems characterized by a different architecture. The equivalence is with respect to a simple input--output function, in this case coin counting, which they multiply realize. 
The systems are also equivalent in terms of their global system dynamics, in the sense that they go through a globally equivalent sequence of internal states. However, because of their different substrates, the three systems specify different cause--effect structures. 
Therefore, based on the postulates of IIT, they are not phenomenally equivalent. In other words, they are equivalent in what they \textit{do} extrinsically, but not in what they \textit{are} intrinsically. 

This dissociation between phenomenal and functional equivalence has important implications. 
As we have seen, a purely feed-forward system necessarily has $\varphi_s = 0$. Therefore, it cannot support a cause--effect structure and cannot be conscious, whereas systems with a recurrent architecture can. 
On the other hand, the behavior (input--output function) of any (discrete) recurrent system can also be implemented by a system with a feed-forward architecture \cite{Krohn1965}. 
This implies that any behavior performed by a conscious system supported by a recurrent architecture can also be performed by an unconscious system, no matter how complex the behavior is. 
More generally, digital computers implementing programs capable of artificial general intelligence may in principle be able to emulate any function performed by conscious humans and yet, because of the way they are physically organized, they would do so without experiencing anything, or at least anything resembling, in quantity and quality, what each of us would experience \cite{Findlay2019}.

The examples also show that the overall system dynamics, while often revealing relevant aspects of a system's architecture, typically do not and cannot exhaust the richness of its current cause--effect structure. For example, a system in a fixed point is dynamically ``dead'' (and ``does'' nothing), but it may be phenomenally quite ``alive,'' for example, experiencing ``pure presence'' \cite{Boly_PurePresence}. 
Of course, the system's causal powers can be fully unfolded, and revealed dynamically, by extensive manipulations and observations of subsets of system units because they are implicitly captured by the system's causal model and ultimately by its transition probability matrix \cite{Albantakis2019cc}.

\subsection*{Conclusions}

IIT attempts to account for the presence and quality of consciousness in physical terms. It starts from the existence of experience, and proceeds by characterizing its essential properties---those that are immediate and irrefutably true of every conceivable experience (axioms). These are then translated into essential properties of physical existence (postulates), the necessary and sufficient conditions that a substrate must satisfy to support an experience---to constitute a complex. ``Physical" is understood purely as cause--effect power, and "substrate'' is meant in purely operational terms---as a set of units that a conscious observer can observe and manipulate.

The postulates can be assessed based purely on a substrate's transition probability matrix, as was illustrated by a few idealized causal models. 
Thus, a substrate of consciousness must be able to take and make a difference upon itself (existence and intrinsicality), it must be able to specify a cause and an effect state that are highly informative and selective (information), and it must do so in a way that is both irreducible (integration) and definite (exclusion). 
Finally, it must specify its cause and effect in a structured manner (composition), where the causal powers of its subsets over its subsets compose a cause--effect structure of distinctions and relations---a $\varPhi$-structure. Thus, a complex does not exist as such but only ``unfolded'' as a $\varPhi$-structure---an \textit{intrinsic entity} that exists for itself, absolutely, rather than relative to an external observer.

As shown above, these requirements constrain what substrates can and cannot support consciousness. Substrates that lack in specificity, due to indeterminism and/or degeneracy, cannot grow to be large complexes. Substrates that are weakly integrated, due to architectural or functional fault lines in their interactions, are less integrated than some of their subsets. Because they are not maximally irreducible, they do not qualify as complexes. This is the case even though they may ``hang together'' well enough from an extrinsic perspective (having a respectable value of $\varphi_s$). Furthermore, even substrates that are maximally integrated may support $\varPhi$-structures that are extremely sparse, as in the case of directed cycles.
Based on the postulates of IIT, a universal substrate ultimately ``condenses'' into a set of disjoint (non-overlapping) complexes, each constituted of a set of macro or micro units.

The physical account of consciousness provided by IIT should be understood as an explanatory identity: every property of an experience should ultimately be accounted for by a property of the cause--effect structure specified by a substrate that satisfies its postulates, with no additional ingredients. 
The identity is not between two different substances or realms---the phenomenal and the physical---but between intrinsic (subjective) existence and extrinsic (objective) existence. Intrinsic existence is immediate and irrefutable, while extrinsic existence is defined operationally as cause--effect power discovered through observation and manipulation. The primacy of intrinsic existence (of experience) in IIT contrasts with standard attempts at accounting for consciousness as something ``generated by'' or ``emerging from'' a substrate constituted of matter and energy and following physical laws.

The physical correspondent of an experience is not the substrate as such but the $\varPhi$-structure specified by the substrate in its current state. Therefore, minor changes in the substrate state can correspond to major changes in the specified $\varPhi$-structure. 
For example, if the state of a single unit changes, an entire $\varPhi$-fold within the $\varPhi$-structure will change, and if a single inactive unit is inactivated, its associated $\varPhi$-fold will collapse, even though the current state of the substrate appears the same (Fig. \ref{fig:state}). 

Each experience corresponds to a $\varPhi$-structure, not a set of functions. Said otherwise, consciousness is about being, not doing \cite{Ellia2021, Tononi_OnBeing, Albantakis2015, Albantakis2019cc}. This means that systems with different architectures may be functionally equivalent---both in terms of global input--output functions and global intrinsic dynamics---but they will not be phenomenally equivalent. For example, a feed-forward system can be functionally equivalent to a recurrent system that constitutes a complex, but feed-forward systems cannot constitute complexes because they do not satisfy maximal irreducibility. Accordingly, artificial systems powered by super-intelligent computer programs would experience nothing (or nearly nothing) because they have the wrong kind of physical architecture, even though they may be behaviorally indistinguishable from human beings \cite{Findlay2019}. 

Even though the entire framework of IIT is based on just a few axioms and postulates, it is not possible in practice to exhaustively apply the postulates to unfold the cause--effect power of realistic systems \cite{Barrett2019, Moyal2020}.
It is not feasible to perform all possible observations and manipulations to fully characterize a universal TPM, or to perform all calculations on the TPM that would be necessary to condense it exhaustively into complexes and unfold their cause--effect power in full. 
The number of possible systems, of system partitions, of candidate distinctions---each with their partitions and relations---is the result of multiple, nested combinatorial explosions. 
Moreover, these observations, manipulations, and calculations would need to be repeated at many different grains, with many rounds of maximizations. For these reasons, a full analysis of complexes and their cause--effect structure can only be performed on idealized systems of a few units \cite{Mayner2018}.

On the other hand, we can simplify the computation considerably by using various assumptions and approximations, as with the ``cut one'' approximation described in \cite{Mayner2018}. Also, while the number of relations vastly exceeds the number of units and of distinctions (its upper bound for a system of $n$ units is $2^{(2^n-1)}-1$), it can be determined analytically, and so can $\sum \varphi_r$ for a given set of distinctions \ref{A:AnalyticalRelations}. Developing tight approximations, as well as bounded estimates of a system's integrated information $\varPhi(s)$, is one of the main areas of ongoing research related to IIT  \cite{Zaeemzadeh}.

Despite the infeasibility of an exhaustive calculation of the relevant quantities and structures for a realistic system, IIT already provides considerable explanatory and predictive power in many real-world situations, making it eminently testable \cite{Tononi2016, Melloni2021, Sarasso2021}. A fundamental prediction is that $\varPhi$ should be high in conscious states, such as wakefulness and dreaming, and low in unconscious states, such as dreamless sleep and anesthesia. 
This prediction has already found substantial support in human studies that have applied a crude approximation of $\varPhi$ to successfully classify subjects as conscious vs. unconscious \cite{Massimini2005, Casarotto2016, Tononi2016, Sarasso2020}. IIT can also account mechanistically for the loss of consciousness in deep sleep and anesthesia \cite{Pigorini2015, Tononi2016}. 
Furthermore, it can provide a principled account of why certain portions of the brain may constitute an ideal substrate of consciousness and others may not, why the borders of the main complex in the brain should be where they are, and why the units of the complex should have a particular grain (the one that yields a maximum of $\varphi_s$). 
A stringent prediction is that the location of the main complex, as determined by the overall maximum of $\varphi_s$ within the brain, should correspond to its location as determined through clinical and experimental evidence.
Another prediction that follows from first principles is that constituents of the main complex can support conscious contents even if they are mostly inactive, but not if they are inactivated \cite{Tononi2016, Haun2019}. 
Yet another prediction is that the complete inactivation of constituents of the main complex should lead to absolute agnosia (unawareness that anything is missing).

IIT further predicts that the quality of experience should be accounted for by the way the $\varPhi$-structure is composed, which in turn depends on the architecture of the substrate specifying it. 
This was demonstrated in a recent paper showing how the fundamental properties of spatial experiences---those that make space feel ``extended''---can be accounted for by those of $\varPhi$-structures specified by 2D grids of units, such as those found in much of posterior cortex \cite{Haun2019}. This prediction is in line with neurological evidence of their role in supporting the experience of space \cite{Haun2019}. 
Ongoing work aims at accounting for the quality of experienced time \cite{Comolatti_Time} and that of experienced objects \cite{Grasso_Objects}. A related prediction is that changes in the strength of connections within the neural substrate of consciousness should be associated with changes in experience, even if neural activity does not change \cite{Song2017}. 
Also, similarities and dissimilarities in the structure of experience should be accounted for by similarities and dissimilarities among $\varPhi$-structures and $\varPhi$-folds specified by the neural substrate of consciousness. Further validation of IIT will depend on a systematic back-and-forth between phenomenology, theoretical inferences, and neuroscientific evidence \cite{Ellia2021}.
 
In addition to empirical work aimed at validating the theory, much remains to be done at the theoretical level. According to IIT, all meaning is intrinsic---whether that of spatial extendedness, of temporal flow, or the feeling of an object that binds a general concept with particular features. 
Every conscious meaning can be considered as identical to a sub-structure within a current $\varPhi$-structure, whether it is triggered by extrinsic inputs or it occurs spontaneously during a dream. 
Ongoing work aims at providing a self-consistent explanation of how intrinsic meanings---whether those of spatial extendedness, of temporal flow, or of objects, to name but a few---can capture relevant features of causal processes in the environment \cite{Mayner_Matching}. It will also be important to explain how intersubjectively validated knowledge can be obtained despite the intrinsic and partially idiosyncratic nature of meaning.  

To the extent that the theory is validated through empirical evidence obtained from the human brain, IIT can then offer a plausible inferential basis for addressing several questions that depend on an explicit theory of consciousness. 
As indicated in the section on phenomenal and functional equivalence, and argued in ongoing work \cite{Findlay2019}, one consequence of IIT is that typical computer architectures are not suitable for supporting consciousness, no matter whether their behavior may resemble ours. 
By the same token, it can be inferred from IIT that animal species that may look and behave quite differently from us may be highly conscious, as long as their brains have a compatible architecture. 
Other inferences concern our own experience and whether it plays a causal role, or is simply ``along for the ride'' while our brain performs its functions. As recently argued, IIT implies that we have true free will---that we have alternatives, make decisions, and cause---and that we, rather than our neurons, are responsible for our actions and their consequences \cite{Tononi2022}.

Finally, an ontology that is grounded in experience as intrinsic existence---an intrinsic ontology---must not only provide an account of subjective existence in objective, operational terms, but also offer a path toward a unified view of nature---of all that exists and happens.
One step in this direction is the application of the same postulates that define causal powers (existence) to the evaluation of actual causes and effects (“what caused what” \cite{Albantakis2019}). 
Another is to unify classical accounts of information (as communication and storage of signals) with IIT's notion of information as derived from the properties of experience---that is, information as causal, intrinsic, specific, maximally irreducible, and structured (meaningful) \cite{Oizumi2014, ZaeemzadehShannon}. 
Yet another is the study of the evolution of a substrate's causal powers as conditional probabilities that update themselves \cite{Albantakis2014}. 

Even so, there are many ways in which IIT may turn out to be inadequate or wrong.
Are some of its assumptions, including those of a discrete, finite set of ``atomic'' units of cause--effect power, incompatible with current physics \cite{Barrett2019, Carroll2021} (but see \cite{Zanardi2018, Kleiner2021_QM, Esteban2018, Kalita2019})? 
Are its axiomatic basis and the translation of axioms into postulates sound and unique? And, most critically, can IIT survive the results of empirical investigations assessing the relationship between the quantity and quality of consciousness and its substrate in the brain?   

\appendix
\section{Appendix}

\subsection{Ties}
\label{A:Ties}

The information postulate requires that a system's cause--effect power is specific: the system in its current state must select a specific cause--effect state for its units. Likewise, mechanisms within the system must select a specific cause and effect state over their purviews. 
The exclusion postulate requires that a complex must be constituted of a definite set of units, and that mechanisms within the complex specify a definite cause and effect. 
By the principle of maximal existence, cause--effect states, complexes, and the cause--effect states of a mechanism within the system are identified as those with maximal cause--effect power.

However, for systems with built-in symmetries in their architecture or the input--output functions of their units, multiple sets of units or cause--effect states may ``tie'' for maximal intrinsic information or cause--effect power \cite{Krohn2017, Moon2019, Hanson2020, Barbosa2021}. Here we outline how those ties should be resolved in line with IIT's postulates and principles.

Maximal substrates can be identified using an iterative algorithm \eqref{eq:sk}. A maximal substrate excludes all overlapping systems with lower $\varphi_s$ from existing as complexes. If overlapping systems tie for $\displaystyle\max_{S\subseteq U_k} \varphi_s(\T_s, s)$ those systems do not comply with the exclusion postulate and we choose the next best system that is unique (see \cite{Marshall2022} for details, and also \cite{Moon2019} for an argument to exclude such ties).

For a system, the maximal cause--effect state $s' = \{s'_c, s'_e\}$ is the one that maximizes the system's intrinsic cause and effect information. If multiple states comply with equation \eqref{eq:s'}, we select the one for which the system specifies the maximal integrated information $\varphi_s(\T_S, s, \theta')$ \eqref{eq:phis} over its minimum partition $\theta'$. Remaining ties in the system state with the same $\varphi_s$ do not matter for system selection, but need to be resolved in order to determine the system's cause--effect structure. By the maximum existence principle, we choose the cause--effect state that maximizes the system's structured information $\varPhi$. In general, any remaining ties in $\varPhi$ in highly symmetrical systems specify cause--effect structures that would be identical from the intrinsic perspective, in which case the tie would be extrinsic and not a violation of the information postulate.

The cause or effect state of a mechanism within the system for a candidate purview is first selected based on its intrinsic information $\ii(m,z)$ \eqref{eq:effect_state}. Next, we compare the integrated information $\varphi(m, Z = z')$ \eqref{eq:phieZ} of all maximal cause or effect states across all possible purviews (including all possible ties in $\ii(m,z)$ within a candidate purview) to identify the maximally irreducible cause or effect $z^*_{c/e}$ of the mechanism within the system \eqref{eq:Zstar}.
By the maximum existence postulate, potential ties in $\max \varphi_d(m, Z = z')$ and thus in the cause--effect state $z^*_{c/e}$ of a distinction may be resolved at the level of the cause--effect structure, by selecting the $z^*_{c/e}$ that maximizes the system's structured information $\varPhi$.
Accordingly, in case of state ties within the same purview, we select the state that is congruent with the system's cause--effect state $s'$. In case of ties across different purviews, the maximal cause--effect state will generally correspond to the one that supports the most relations with other distinctions, which typically favors larger purviews. 

\subsection{Comparison to IIT 1.0--3.0 and subsequent publications}
\label{A:Comparison}
As highlighted in the main text, IIT is a work in progress. While the core theory has remained the same, its formal framework has been progressively refined and extended \cite{Tononi2003, Tononi2004, Balduzzi2008, Oizumi2014}. Compared to prior versions (IIT 1.0 \cite{Tononi2003, Tononi2004}, IIT 2.0 \cite{Balduzzi2008, Balduzzi2009, Tononi2008}, and IIT 3.0 \cite{Tononi2012, Oizumi2014, Mayner2018, Kleiner2021_QM}), IIT 4.0 presents a more complete, self-consistent formulation.
The most notable advances in IIT 4.0 include the introduction of an Intrinsic Difference (ID) measure \cite{Barbosa2020, Barbosa2021} that is uniquely consistent with IIT's postulates, the explicit assessment of causal relations \cite{Haun2019}, and a more exact translation of the axioms into postulates. 
Because IIT 3.0 already included a comparison to IIT 1.0 and 2.0 (see \cite{Oizumi2014}, Supporting Information Text S1), we will mainly focus on subsequent developments.

\subsubsection*{Axioms and postulates}
While the starting point of IIT has always been phenomenology, the axioms and postulates of the theory were first explicitly presented in IIT 3.0 \cite{Oizumi2014, Tononi2016}. 
The updated 4.0 exposition of IIT's axioms explicitly separates phenomenal existence, which is not a property, from intrinsicality, which is one of the essential properties of phenomenal existence. Accordingly, the existence of experience is introduced as IIT's foundational, zeroth axiom. The remaining five axioms (intrinsicality, information, integration, exclusion, and composition) capture the essential properties that are immediate and irrefutably true of every conceivable experience.

Compared to IIT 3.0, the formulation of the axioms has been refined to avoid misunderstandings \cite{Bayne2018, Merker2021} and to highlight their immediacy and irrefutability. The formulation of IIT's postulates has been updated accordingly with the objective of tracking the phenomenal axioms as closely as possible.
For example, conforming more closely to the information axiom, the information postulate requires that the system must select a specific cause--effect state over its units.
The composition axiom now highlights both phenomenal distinctions and their relations. 
By the composition postulate, phenomenal distinctions and relations are accounted for in physical terms by causal distinctions and relations. 
Because only an experience that exists intrinsically, in a way that is specific, irreducible, and definite can also be structured, composition takes the final position in the ordering of the axioms and postulates.

\subsubsection*{Identifying maximal substrates}
The IIT 4.0 formalism to identify maximal substrates was first described in detail in \cite{Marshall2022}.
Maximal substrates, or complexes, are identified based on their system integrated information $\varphi_s$ (as in IIT 2.0 but unlike IIT 3.0, which evaluated integration after composition). 
System partitions remain directional (as in IIT 3.0). In IIT 4.0, the minimum partition (MIP) is identified as the partition with minimal integrated information ($\varphi_s$), normalized by the maximal possible value of $\varphi_s$ across this partition for an arbitrary TPM of the same dimensions \eqref{eq:mips}. In this way, the MIP is sensitive to the fault lines of a candidate system, rather than defaulting to partitions of individual system units.
The IIT 4.0 analysis is state-dependent (as in IIT 2.0 and 3.0) and requires positive cause and effect power for a system to exist (as in IIT 1.0 and 3.0). 

\subsubsection*{Measuring intrinsic information}
Supplanting prior measures such as the Kullback-Leibler divergence (IIT 2.0 \cite{Balduzzi2008}), or the (extended) Earth Mover's Distance (EMD, IIT 3.0, \cite{Oizumi2014}), the IIT 4.0 formalism features a newly developed Intrinsic Difference (ID) measure \cite{Barbosa2020}, which uniquely complies with the postulates of IIT.
Accordingly, the intrinsic information specified by a system (or mechanism) over a cause or effect state is evaluated as a product of informativeness and selectivity, which makes it subadditive if $p < 1$, that is, if cause--effect power is spread over more than one state.
As intrinsic information is evaluated over specific cause or effect states, its maximal value over a state distribution identifies the specific cause or effect state selected by the system (or mechanism), in line with the information postulate.

\subsubsection*{Causal distinctions}
Distinctions capture how the various system subsets specify system subsets as their cause and effect within the system. 
In IIT 3.0, distinctions were called ``concepts'' (which composed to a ``conceptual'' structure), a term that could generate unnecessary misunderstandings \cite{Haun2019}. 
An updated formalism to identify causal distinctions was first presented in \cite{Barbosa2021}.
As in IIT 3.0, causes and effects must be specified in a way that is irreducible ($\varphi_d > 0$). Unlike IIT 3.0, in IIT 4.0 distinctions select a specific state as a cause and an effect. 
The definition of cause and effect probabilities $\pi(z\mid m)$ (\eqref{eq:cause_probability} and \eqref{eq:effect_probability}) remains unchanged, but is now presented more formally in terms of product probabilities, rather than referring to ``virtual elements'' \cite{Krohn2017, Albantakis2019}). 
In IIT 4.0, a mechanism selects a specific cause and effect state based on the Intrinsic Difference (ID) measure introduced in \cite{Barbosa2020} (see above). The set of permissible partitions $\theta \in \Theta(M,Z)$ \eqref{eq:partitions} has also been updated \cite{Albantakis2019, Barbosa2021} to ensure that partitions are always ``disintegrating'' the mechanism.

The present formulation includes several updates compared to \cite{Barbosa2021}. First, the cause--effect state $z'$ is selected based on the intrinsic information $\ii_e(m,Z)$ \eqref{eq:effect_state}, before evaluating its integrated information $\varphi(m,Z = z')$ \eqref{eq:phieZ}. This is because the cause/effect of $m$ should be determined by the mechanism as a whole, independent of how it can be partitioned. 
Second, we evaluate intrinsic information without the absolute value, as in \cite{Barbosa2020}, because, to comply with the existence postulate, a mechanism's cause state should be one that would increase the probability of its current state, and its effect state one whose probability would be increased by the mechanism being in its current state. 
Third, to correctly capture this increase in probability on the cause side, the informativeness term is expressed in terms of effect probabilities also on the cause side for $\ii_c$ and $\varphi_c$, evaluating the increase in probability of the current state due to the cause state.
Fourth, we have updated the resolution of ties at the level of distinctions according to the principle of maximal existence (see \ref{A:Ties}).
Finally, a candidate distinction only contributes to the system's cause--effect structure if its maximal cause--effect state $z^*$ is congruent with the maximal cause--effect state $s'$ of the system.

\subsubsection*{Relations}
Relations bind together a set of causal distinctions over a congruent overlap between their causes and/or effects. 
Developing an explicit account of phenomenal relations in terms of causal relations was a main goal of IIT since IIT 1.0. IIT 3.0 employed a distance metric---the Earth Mover's Distance---that was sensitive to whether different distinctions (``concepts'') had similar cause--effect repertoires, but relations did not figure explicitly in the formalism despite their central role in characterizing experience. 
An explicit account was first described in \cite{Haun2019}.
The IIT 4.0 formalism further distinguishes between relations (which bind a set of distinctions with overlapping causes and/or effects) and the faces of a relation (which specify the maximal overlap of a set of purviews and jointly characterize the type of the relation). 
Moreover, the amount of information specified by a distinction over the overlap and the way relation partitions are assessed differs from the original account \cite{Haun2019}. Because distinctions are irreducible components within the cause--effect structure upon which relations are built, a distinction involved in a relation contributes its entire $\varphi_d$, weighted by the extent of its joint overlap \eqref{eq:phir}. For this reason, we do not recompute the irreducible information of the mechanism $m$ of a distinction $d(m, z^*, \varphi_d)$ over the candidate overlap $o$. In \cite{Haun2019}, distinctions contributing to a relation were partitioned by ``noising'' the interactions among distinction units. 
(For the same reason, once the system is ``unfolded'' and its candidate distinctions are established, the probabilities of $z^* \in \bm{z}$ are no longer referenced). 

\subsubsection*{$\varPhi$-structures}
In IIT 4.0, a system is a substrate of consciousness (a complex) if it corresponds to a maximum of system integrated information $\varphi_s$, as determined through information, integration, and exclusion. 
This is similar to IIT 2.0 \cite{Balduzzi2008}, although in that case only causes (and not effects) were evaluated. 
The quality of the experience is identical to the $\varPhi$-structure of distinctions and relations unfolded from the complex. 
The quantity of experience corresponds to the $\varPhi$ value, the sum of the integrated information of the distinctions and relations that compose the $\varPhi$-structure. 
In IIT 3.0, the determination of the complex through information, integration, and exclusion took into account its compositional structure, although without explicit relations. However, the $\varPhi$ value corresponding to the quantity of consciousness only captured the distinctions affected by the minimum partition, as opposed to all the distinction (and relations) unfolded from a maximally irreducible substrate. 

\subsection{Analytical solution for $\sum \varphi_r$ and the number of causal relations}
\label{A:AnalyticalRelations}
Here, we show how the sum of the relation integrated information over all the causal relations ($\sum \varphi_r$) and the number of relations can be computed without assessing the relations individually. We only need the set of causal distinctions:

$$
    D(T_S, s) = \{d(m) ~:~m \subseteq s, ~ \varphi_d(m) > 0, ~z^*_c(m) \subseteq s'_c, ~z^*_e(m) \subseteq s'_e \},  
$$
where $d(m) = (m, z^*(m), \varphi_d(m) )$ and $
z^*(m) = \{ z^*_c(m), z^*_e(m) \}$. 

\subsubsection*{Analytical computation of $\sum \varphi_r$}
Given a subset of distinctions $\bm{d} \subseteq D(T_S, s)$ with $|\bm{d}| \geq 2$, any subset $\bm{z}$ of purviews that contains either the cause, or effect, or both the cause and effect of each distinction $d \in \bm{d}$ and overlap congruently defines a relation face $f$ with face overlap $o_f^* = \bigcap_{z \in \bm{z}} z$. The relation overlap is further defined as the union of the face overlaps $\bigcup_{f \in \bm{f}(\bm{d})} o_f^*$, where $\bm{f}(\bm{d})$ represents the set of all the faces over the distinction set $\bm{d}$. Here, intersection and union take into account both the units and their states. 

First, we can show:

$$
\bigcup_{f \in \bm{f}(\bm{d})} o_f^* = \bigcap_{d \in \bm{d}} \big(z^*_c(d) \cup z^*_e(d)\big),
$$
by proving any unit $n$ in $\bigcup_{f \in \bm{f}(\bm{d})} o_f^*$ is in $\bigcap_{d \in \bm{d}} 
\big(z^*_c(d) \cup z^*_e(d)\big)$ and vice versa:

$$
\begin{aligned}
n \in \bigcup_{f \in \bm{f}(\bm{d})} o_f^* 
\iff \exists f \in \bm{f}(\bm{d}), n \in o_f^* 
\iff \forall d \in \bm{d}, n \in z^*_c(d) \ \text{or} \ n \in z^*_e(d) \\
\iff \forall d \in \bm{d}, n \in z^*_c(d) \cup z^*_e(d)
\iff n \in \bigcap_{d \in \bm{d}} \big(z^*_c(d) \cup z^*_e(d)\big)
\end{aligned}
$$
This helps us to rewrite the relation integrated information of a set of distinctions $\bm{d} \subseteq D(T_S, s)$ with $|\bm{d}| \geq 2$ as:

$$
\left|\bigcap_{d \in \bm{d}} \big(z^*_c(d) \cup z^*_e(d)\big) \right | \min_{(z_d, \varphi_d) \in \bm{d}} \frac{\varphi_d}{|z^*_c(d) \cup z^*_e(d)|}.
$$
We further define the set of $z^*_c(d) \cup z^*_e(d)$ of all distinctions in $D$ and their corresponding distinction integrated information as:

$$
\mathcal{Z}(T_S, s) = \{ \big( z^*_c(m) \cup z^*_e(m), \varphi(m) \big) : (m, z^*(m), \varphi_d(m) ) \in D(T_S, s) \}.
$$
Now, given a single node $n$ in a specific state, we can find all the distinctions that contain $n$ in that state in their cause, or effect, or both purviews as:

\begin{equation}
\label{eq:purview_inclusion_single_node}
\mathcal{Z}(n) = \{ (z, \varphi) : (z, \varphi) \in \mathcal{Z}(T_S, s), n \in z   \}.
\end{equation}

Any subset of $\mathcal{Z}(n)$ of size $2$ or larger defines a relation whose overlap contains at least $n$. Formally, for $\bm{r} \subseteq \mathcal{Z}(n)$, $|\bm{r}| \geq 2$, there exists a relation with relation purview $\bigcap_{(z_d, \varphi_d) \in \bm{r}} z_d$ and integrated information value of:

$$
\left|\bigcap_{(z_d, \varphi_d) \in \bm{r}} z_d \right | \min_{(z_d, \varphi_d) \in \bm{r}} \frac{\varphi_d}{|z_d|}.
$$
Note that, by definition of $\mathcal{Z}(n)$ and $\mathcal{Z}(T_S, s)$, $z_d$ is the union of cause and effect purviews.
Using the definition of $\mathcal{Z}(T_S, s)$ and $\mathcal{Z}(n)$, we can write the sum of the integrated information of relations, except self-relations, as 

$$
\sum_{\substack{\bm{r} \subseteq \mathcal{Z}(T_S, s)\\
                  \bm{r} \geq 2}}
                  \left|\bigcap_{(z_d, \varphi_d) \in \bm{r}} z_d \right | \min_{(z_d, \varphi_d) \in \bm{r}} \frac{\varphi_d}{|z_d|} = 
\sum_{n \in s'_c \cup s'_e}\sum_{\substack{\bm{r} \subseteq \mathcal{Z}(n)\\
                  |\bm{r}| \geq 2}} \min_{(z_d, \varphi_d) \in \bm{r}} \frac{\varphi_d}{|z_d|},
                  $$
By factoring the sum over $\bm{r} \subseteq \mathcal{Z}(T_S, s)$ into two sums over the nodes $n$ and the relations whose purview contains \emph{at least} $n$, $\bm{r} \subseteq \mathcal{Z}(n), |\bm{r}| \geq 2$, we are overcounting each relation by a factor of its joint purview size $\left|\bigcap_{(z_d, \varphi_d) \in \bm{r}} z_d \right |$. For example, if a set of distinctions make up a relation $\bm{r}$ over two units $n_1$ and $n_2$, they all are members of both $\mathcal{Z}(n_1)$ and $\mathcal{Z}(n_2)$. Therefore, $\bm{r} \subseteq \mathcal{Z}(n_1)$ and $\bm{r} \subseteq \mathcal{Z}(n_2)$. This simplifies the summand to just $\min_{(z_d, \varphi_d) \in \bm{r}} \frac{\varphi_d}{|z_d|}$. To compute the inner sum, we can sort the distinctions in $\mathcal{Z}(n)$ by their $\frac{\varphi}{|z|}$ value in a non-decreasing order, such that $(z_{(1)}, \varphi_{(1)})$ has the smallest $\frac{\varphi}{|z|}$ ratio, $(z_{(2)}, \varphi_{(2)})$ has the second smallest $\frac{\varphi}{|z|}$ ratio, and so on. Then, we can compute the sum as:

$$
\sum_{\substack{\bm{r} \subseteq \mathcal{Z}(n)\\
                  |\bm{r}| \geq 2}} \min_{(z_d, \varphi_d) \in \bm{r}} \frac{\varphi_d}{|z_d|} =
\sum_{j = 1}^{|\mathcal{Z}(n)|} \frac{\varphi_{(j)}}{|z_{(j)}|} (2^{|\mathcal{Z}(n)| - j} - 1).
$$

In words, any subset $\bm{r} \subseteq \mathcal{Z}(n), |\bm{r}| \geq 2$, that contains $(z_{(1)}, \varphi_{(1)})$ will have $\min_{(z_d, \varphi_d) \in \bm{r}} \frac{\varphi_d}{|z_d|} = \frac{\varphi_{(1)}}{|z_{(1)}|}$. There are $2^{|\mathcal{Z}(n)| - 1} - 1$ of such subsets. Similarly, there 
are $2^{|\mathcal{Z}(n)| - 2} - 1$ subsets that contain $(z_{(2)}, \varphi_{(2)})$, but not $(z_{(1)}, \varphi_{(1)})$, etc. This helps us arrive at our final results:

$$
\sum_{\substack{\bm{r} \subseteq \mathcal{Z}(T_S, s)\\
                  |\bm{r}| \geq 2}}
                  \left|\bigcap_{(z_d, \varphi_d) \in \bm{r}} z_d \right | \min_{(z_d, \varphi_d) \in \bm{r}} \frac{\varphi_d}{|z_d|}
= 
\sum_{n \in s'_c \cup s'_e}
\sum_{j = 1}^{|\mathcal{Z}(n)|} \frac{\varphi_{(j)}}{|z_{(j)}|} (2^{|\mathcal{Z}(n)| - j} - 1).$$
This gives us the sum of the relation integrated information of all the relations, except the self-relations, i.e. $|\bm{r}| = 1$. The self-relations can be assessed individually without combinatorial explosion. 

\subsubsection*{Analytical count of the number of relations}
We can also count all the causal relations among all the distinctions in $D(T_S, s)$ by generalizing the definition of $\mathcal{Z}(n)$ in \eqref{eq:purview_inclusion_single_node} to all the subsets $o \subseteq s'_c \cup s'_e$:
$$
\mathcal{Z}(o) = \{ (z, \varphi) : (z, \varphi) \in \mathcal{Z}(T_S, s), z \supseteq o   \}.
$$
For each distinction $d \in D(T_S, s)$, there is a corresponding element $(\big(z^*_c(d) \cup z^*_e(d), \varphi(d) \big)$ in $\mathcal{Z}(o)$ if $o \subseteq z^*_c(d) \cup z^*_e(d)$. Any subset of $\mathcal{Z}(o)$ of size $2$ or larger defines a relation whose overlap contains at least $o$. The number of such subsets is:
$$
2^{|\mathcal{Z}(o)|} - |\mathcal{Z}(o)| - 1.
$$
We can count all the relations by applying the inclusion-exclusion principle as:
$$
\sum_{o \subseteq s'_c \cup s'_e} (-1)^{|o|-1} \left( 2^{|\mathcal{Z}(o)|} - |\mathcal{Z}(o)| - 1 \right).
$$
This is the number of all the causal relations among the causal distinctions in $D(T_S, s)$, except the self-relations. Again, the self-relations can be counted individually without combinatorial explosion.

\subsection{IIT Algorithm}
\label{A:Algorithm}
\begin{figure}[p]
\hspace*{-4cm} 
    \includegraphics[width=\columnwidth+4cm]{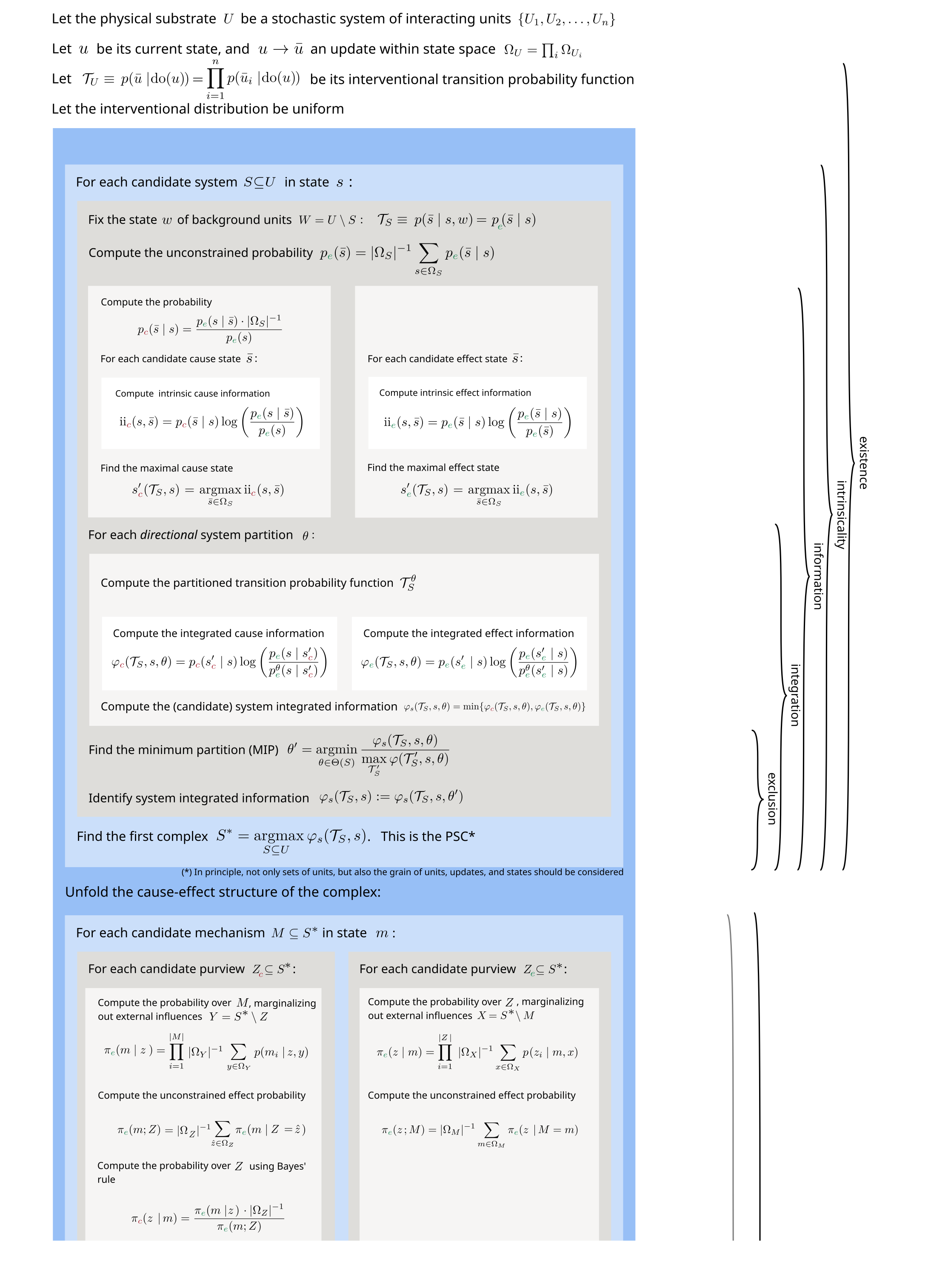}
\end{figure}
\begin{figure}[p]
\vspace*{-1cm}
\hspace*{-4cm} 
    \includegraphics[width=\columnwidth+4cm]{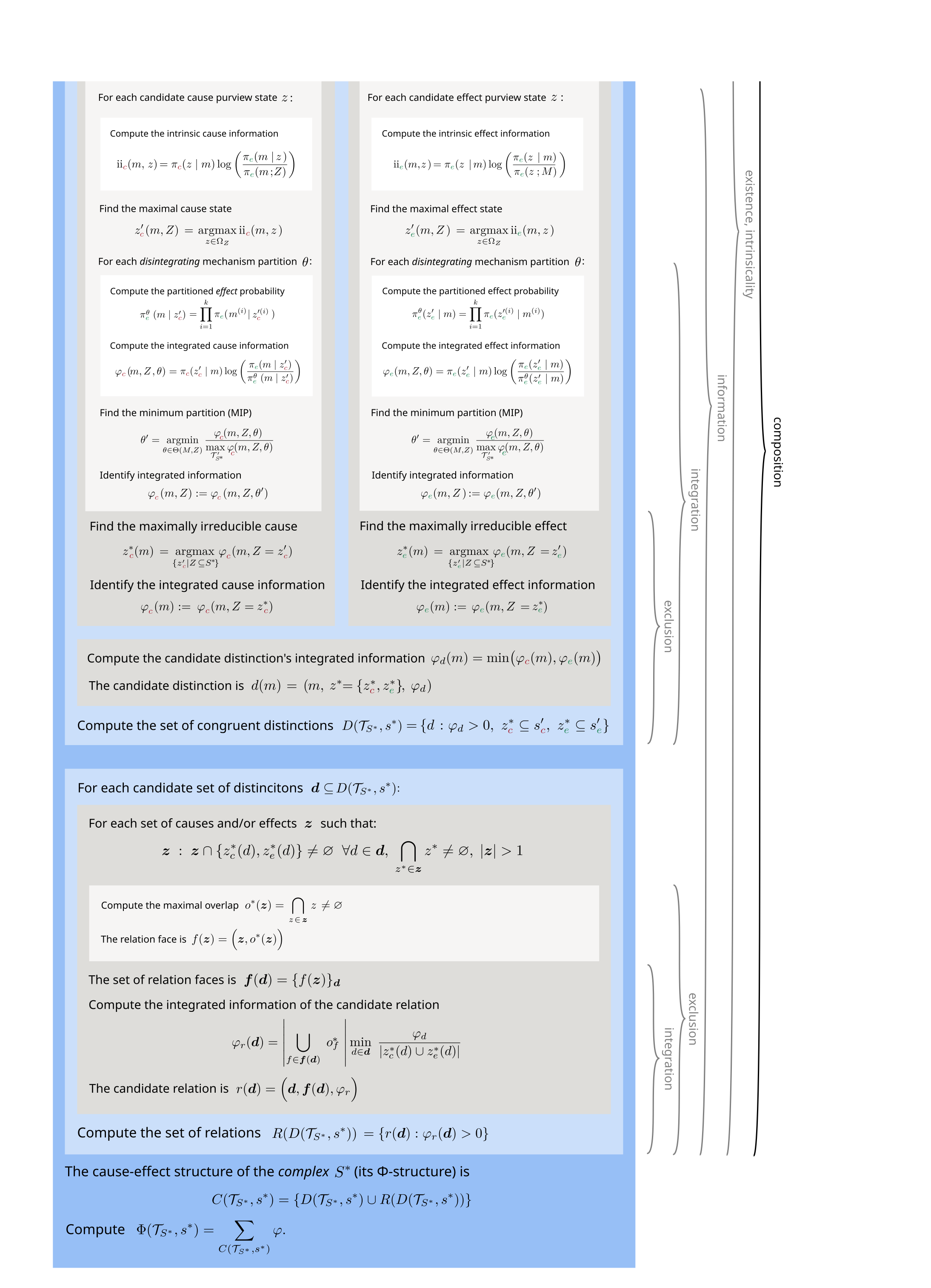}
\end{figure}
\newpage

\section*{Acknowledgments}
This project was made possible through the support of a grant from Templeton World Charity Foundation (TWCF0216). In addition, this research was supported by the David P White Chair in Sleep Medicine at the University of Wisconsin-Madison, by the Tiny Blue Dot Foundation (UW 133AAG3451; G.T.), and by the Natural Science and Engineering Research Council of Canada (NSERC; RGPIN-2019-05418; W.M.). L.A. also acknowledges the support of a grant from the Templeton World Charity Foundation (TWCF-2020-20526).

\nolinenumbers

\bibliography{IIT_bibtex.bib}

%
%
%





\end{document}